\documentclass[draftcls,onecolumn]{IEEEtran}
\ifCLASSINFOpdf
\else
\fi
\usepackage{color}
\usepackage{cite}
\usepackage[cmex10]{amsmath}
\usepackage{amsthm}
\usepackage{algorithm}
\usepackage{algorithmic}
\usepackage{stfloats}
\usepackage{subfigure}
\usepackage{mathrsfs}
\usepackage{multicol,multienum}
\usepackage[pdftex]{graphicx}

\begin{document}
%

\title{Power Control in UAV-Supported Ultra Dense Networks: Communications, Caching, and Energy Transfer}

\author{
        {
        Haichao Wang,
        Guoru Ding,
        Feifei Gao,
        Jin Chen,
        Jinlong Wang,
        and
       Le Wang
       }

\thanks{\emph{Haichao Wang, Guoru Ding (corresponding author), Jin Chen, Jinlong Wang, and Le Wang are with PLA University of Science and Technology; Guoru Ding is also with Southeast University; Feifei Gao is with Tsinghua University}}
}

%
%

\maketitle
\begin{abstract}
By means of network densification, ultra dense networks (UDNs) can efficiently broaden the network coverage and enhance the system throughput. In parallel, unmanned aerial vehicles (UAVs) communications and networking have attracted increasing attention recently due to their high agility and numerous applications. In this article, we present a vision of UAV-supported UDNs. Firstly, we present four representative scenarios to show the broad applications of UAV-supported UDNs in communications, caching and energy transfer. Then, we highlight the efficient power control in UAV-supported UDNs by discussing the main design considerations and methods in a comprehensive manner. Furthermore, we demonstrate the performance superiority of UAV-supported UDNs via case study simulations, compared to traditional fixed infrastructure based networks. In addition, we discuss the dominating technical challenges and open issues ahead.
\end{abstract}


%
\IEEEpeerreviewmaketitle

\section*{Introduction}
\IEEEPARstart{T}{he} quick development of the mobile Internet and Internet of Things (IoT) brings critical challenges to the design of mobile wireless networks mainly for providing ultra high data rate and very low time delay. According to a recent ITU report\cite{ITU}, the wireless data traffic will be 10,000 times in 2030 compared to that in 2010. Ultra dense network (UDN) is a favorable technique to accomplish the requirements for explosive data traffic\cite{UDNMag1}. Thanks to the flexible deployment and low transmit power, deploying massive smallcell base stations (SBSs) can efficiently broaden the network coverage and enhance the overall throughpu\cite{UDNser}. Most of existing studies on UDNs have focused on improving the performance of terrestrial heterogeneous cellular networks by addressing various issues, such as, the coexistence of WiFi and heterogeneous smallcell networks\cite{ZHJ}, user association and resource allocation\cite{ZJC}, and energy/spectral efficient frequency reuse in heterogeneous ultra dense scenarios\cite{EE_LYS,EE_MD}, to name just a few.

In parallel, unmanned aerial vehicles (UAVs) communications and networking have attracted increasing attention recently due to their high agility and numerous applications. Introducing UAVs into UDNs can achieve significant gains by fully exploiting their potentials\cite{ZR}. UAVs can be rapidly deployed to serve the wireless users without being hampered by geographical constraints compared to traditional terrestrial infrastructure. For example, they can act as flying base stations (BSs) to enhance wireless coverage and boost throughput at hotspots such as campuses and sport stadiums, or in the region where the cellular infrastructure is unavailable. They can also act as flying/mobile relays at these areas where the communications appear among separated users without reliable direct communication links. In addition, UAVs can achieve effective relocation in response to the users' mobility. By dynamically adjusting the locations of UAVs, one can establish almost line-of-sight (LOS) communication links in most scenarios, thereby significantly improving the system performance.

To exploit the potential merits, this article presents a vision of UAV-supported UDNs and investigates the power control problem in UAV-supported UDNs. Firstly, we present four representative scenarios to show the broad applications of UAV-supported UDNs in communications, caching and energy transfer. Then, we highlight the efficient power control in UAV-supported UDNs by discussing the main design considerations and methods in a comprehensive manner. Furthermore, we demonstrate the performance superiority of UAV-supported UDNs via case study simulations, compared to traditional fixed infrastructure based networks. In addition, we discuss the dominating technical challenges and open issues ahead.
\section*{UAV-Supported UDNs}
\begin{figure}[!t]
\centering{\includegraphics[width=140mm]{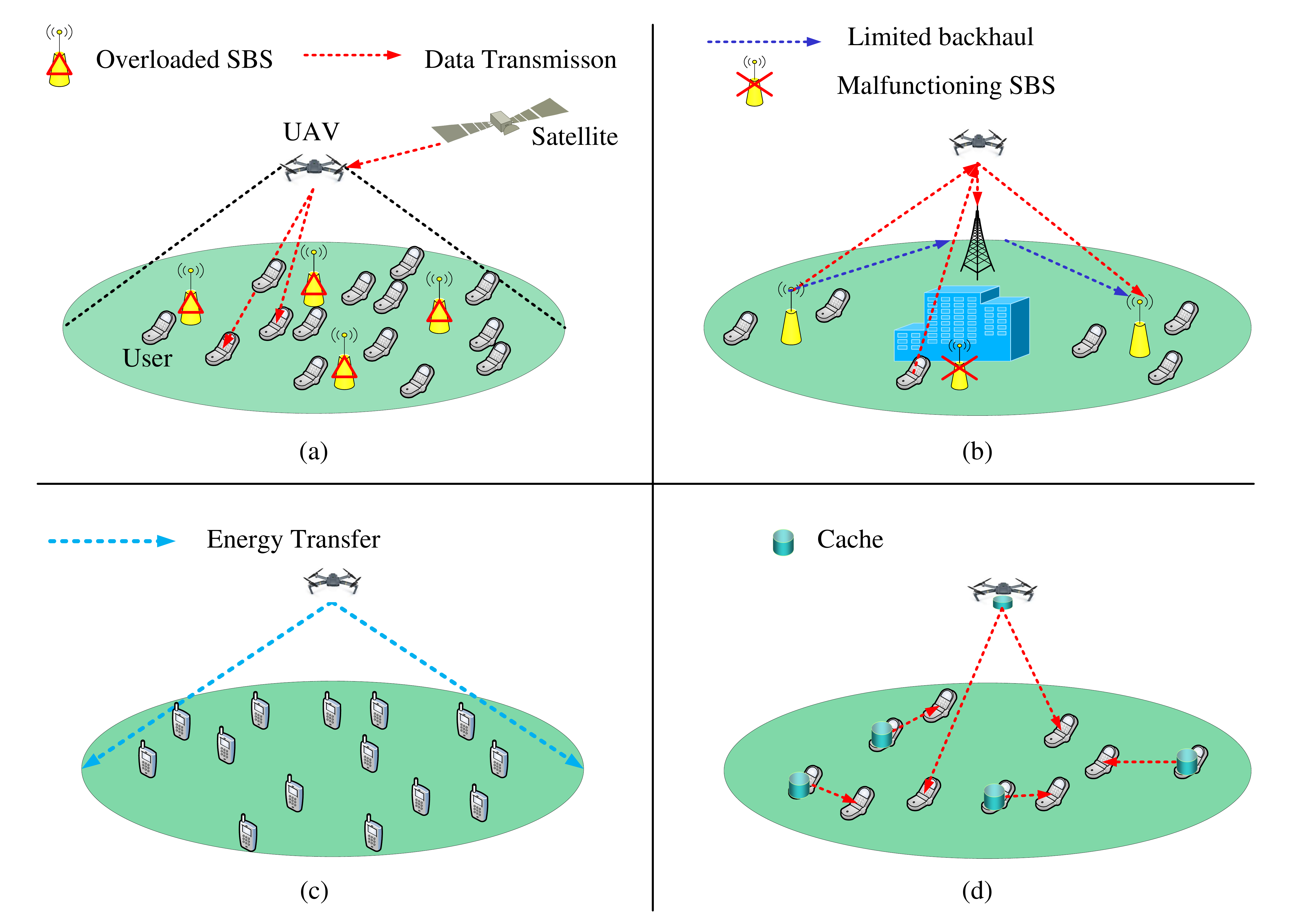}}
\caption{Four representative cases for UAV-supported UDNs: a) UAV-supported BS; b) UAV-supported relay; c) UAV-supported energy transfer; d) UAV-supported caching.}
\label{model}
\end{figure}

In contrast to the conventional terrestrial infrastructure with fixed location, UAVs have unparalleled advantages due to their inherent mobility. UDNs will usher in major developments and changes with UAV supporting. Figure. \ref{model} shows four representative scenarios with UAV supporting.
\begin{itemize}
  \item \emph{UAV as flying BSs:} UAVs acting as flying BSs provide connectivity for the users in overloaded cells or the cases without infrastructure for wireless access, such as the regions where the cellular infrastructure has been damaged due to natural disasters.
  \item \emph{UAV as mobile relays:} UAVs as mobile relays can cooperate to forward the information among ground stations with limited backhauls or the separated nodes with negligible direct links.
  \item \emph{UAV as motorial energy sources:} UAVs can carry an energy source or even act as an energy source to charge the wireless nodes for prolonging the network lifetime. Typical applications include wireless sensor networks and IoT where wire charging is unavailable.
  \item \emph{UAV as aerial caches:} UAVs can be employed as caches to effectively deliver the contents by tracking the corresponding destinations. One typical application is delay-tolerant surveillance in agriculture, soil, and ocean.
\end{itemize}

There are many types of UAVs that can be used in the application scenarios above. In this article, we mainly consider medium and large UAVs with greater load capacity and flight duration.
\subsection*{UAV-Supported BS}
The spatial utilization of limited spectrum can be greatly improved by deploying UDNs, enabling more users to be served simultaneously. Even so, the traffic may be still overloaded in some hotspots at certain time (e.g., sport stadiums, attendees of festivals, or campuses). To offload the wireless data traffic of user equipments from macrocells or smallcells, UAVs can act as flying BSs to rapidly provide wireless access for ground users\cite{UAV-BS,MM2}. The optimal altitude of the UAV resulting in the maximum sum-rate gain is investigated in\cite{UAV-BS}, where a UAV serves a number of ground nodes within its coverage area. In\cite{MM2}, the authors analyze the coverage and rate performance when a UAV is employed as a flying BS. Google and Facebook both seek to build UAV communications networks, which are independent of the existing Internet service providers. For UAV-supported BS, a certain amount of time-frequency resource should be assigned for UAVs as backhauls. Moreover, for multi-UAVs communications, some UAVs acts as sink nodes to communicate with the ground control center, which require better backhauls. Since the total resource is limited, how to allocate these resource should be carefully designed.
\subsection*{UAV-Supported Relay}
In UDN environments, as the number of smallcells increases, it might be challenging for operators to preserve an ideal backhaul for each SBS\cite{UDNser}. Moreover, some SBSs are deployed in hard-to-reach places, and thus ideal backhaul links are more difficult to be achieved. Furthermore, any communication infrastructure may be interrupted due to natural disasters or man-made destruction, which results that the users they serve lose network connections. For these cases, UAVs can be quickly and swiftly deployed to assist the existing communication infrastructure. Authors in\cite{relaytc} study the throughput maximization problem in a mobile relay system by optimizing the source/relay transmit power along with the relay trajectory. It has been shown that UAVs can achieve considerable throughput gain by dynamically adjusting their locations as opposed to static relays.
\subsection*{UAV-Supported Energy Transfer}
UDNs generally consist of massive spatially distributed wireless nodes, e.g., device-to-device (D2D) communications, machine-to-machine communications or sensor nodes. One challenging problem for these networks is to reduce the energy consumption and prolong the network lifetime since batteries are the primary energy sources. Regularly recharging or replacement of batteries for massive nodes can be costly and inconvenient. Wireless energy harvesting and transfer has been introduced as a promising enabler to promote green communications. UAV-based wireless energy transfer is a very promising solution since the UAV is able to cover a large area in a relatively short time. Moreover, the distance between the energy source and destination can be artificially and adaptively adjusted to improve the energy harvesting efficiency and satisfy the needs of different nodes. Notice that the UAV can carry an energy source and thus not consume its energy to transfer wireless energy. In this sense, medium UAVs with smaller sizes are much more efficient than large UAVs.
\subsection*{UAV-Supported Caching}
Although human behavior is difficult to be predicted accurately, many users will request the same popular contents at different time. By ahead of time caching the most popular contents in local memory, wireless caching technology can relieve the data access pressure in UDNs\cite{M2M}. However, traditional static caching devices do not meet the users' mobility requirements. The inherent mobility feature of UAVs can be fully utilized to realize this vision. Interestingly, authors in\cite{Cache} have derived the optimal locations of UAVs as well as the content to cache for cache-enabled UAVs. Besides, the UAV-supported caching plays an important role in environmental monitoring, such as agriculture, soil, and ocean. They can cache the information collected by sensors and transmit to the ground center within the tolerant delay.
\section*{Efficient Power Control in UAV-Supported UDNs}
To fully exploit the great potentials of UAVs, power control strategy should be carefully designed, which requires a deep change from terrestrial networks toward aerial networks, from fixed infrastructure based networks toward mobile infrastructure based networks. In this context, except for the conventional time-frequency resource, the UAV's trajectory, speed, number, location and energy should be considered to assist the power control.
\subsection*{Main Design Considerations}
From the system perspective, the aim of power control in UAV-supported UDNs is to maximize or minimize one specific objective while satisfying given constraints, which are discussed in the following.

\textbf{Coverage/outage probability:} Coverage and outage probability are important performance metrics when UAVs are used to provide wireless connectivity for mobile users, e.g., acting as flying BSs or relays. In UAV-supported UDNs, the mutual interference among massive UAVs or UAVs and SBSs is serious and thus it cannot be ignored. Increasing the transmit power can improve the coverage probability of certain access point. But this at the same time makes other users suffer more serious interference, resulting in the increase of outage probability.

\textbf{Spectral/energy efficiency:} Spectral and energy efficiency reflect the resource utilization from different perspectives. While deploying massive UAVs improves the system spectral efficiency mainly via spatial reuse, some characteristics of UAVs should also be taken into account in UAV-supported UDNs. That is, limited available energy is still restricting the wide applications of UAVs despite the fact that energy storage technology has made great progress. Therefore, energy-aware optimization is crucial for UAV-supported UDNs.

\textbf{Network rate/delay:} In UAV-supported UDNs, e.g., when UAVs act as mobile caches for delay-tolerant surveillance in agriculture, soil, and ocean, high data rate is pursued. Meanwhile, the delay is also expected to be small. However, there exists a tradeoff between the network rate and delay. In particular, high rate may account for high latency and vice versa. Generally, given a sufficiently high delay tolerance, UAV-supported UDNs can realize better rate performance.

\subsection*{Power Control Methods}
Generally speaking, the power control methods can be divided into centralized and decentralized methods from the network architecture point of view.

\begin{figure}[!t]
\centering{\includegraphics[width=120mm]{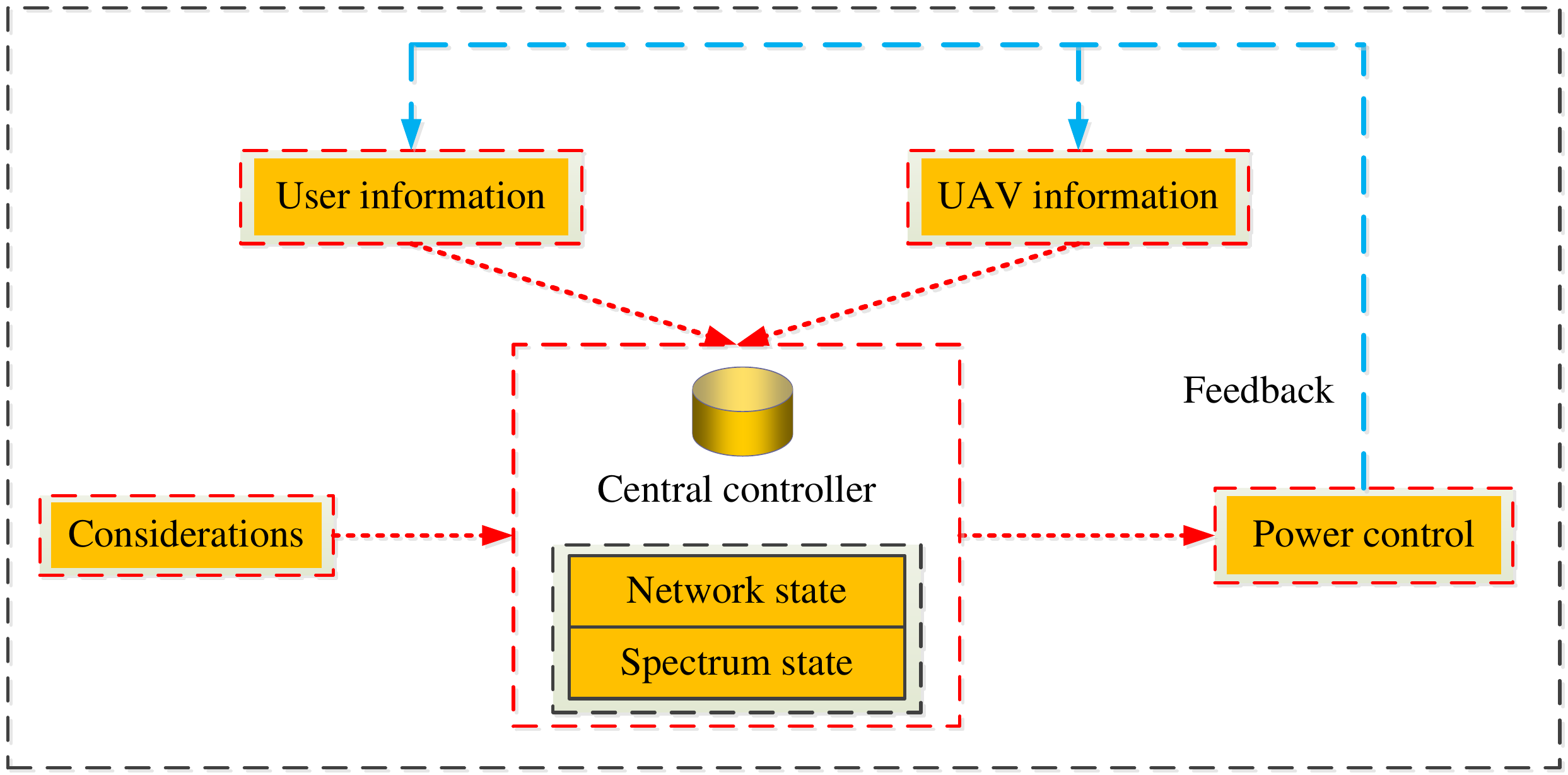}}
\caption{The centralized power control architecture for UAV-supported UDNs.}
\label{Centrilized}
\end{figure}

\textbf{Centralized approaches:} Centralized methods fit the case that the network state and channel state information (CSI) change slowly. A central controller is utilized to gather all the information and further make decisions tailored to the network and spectrum state as shown in Fig. \ref{Centrilized}. The network state is related to the current network deployment of other access points (femtocells, picocells and WiFi, etc.), which will interact with UAV networks. The spectrum state refers to the available spectrum and available time. User information includes its demand, type, location, mobility and energy. UAV information mainly refers to its location, speed, cached content, processing capacity and stored energy, etc. UAVs or users perform appropriate operations after receiving the power control strategy broadcasted by the central controller.

There are several frequently-used centralized optimization methods, such as convex optimization, fractional programming, geometric programming, and successive convex approximation, etc. They can get an optimal solution in some special cases. However, finding an optimal solution is not always tractable in UAV-supported UDNs. On the one hand, the UAV's trajectory is continuous, meaning that we need to solve a problem with infinite variable. On the other hand, the complexity of centralized approaches dramatically grows with increasing number of users, which is not what we wished since this would lead to high delay especially in UDNs. Therefore, a suboptimal or an acceptable solution is expected to be obtained in a short period of time.

\textbf{Decentralized approaches:} Although the centralized approaches can realize better solutions, there must be a central controller collecting, storing, and distributing the global information. These kinds of methods make the system subject to heavy signaling overhead and result in poor scalability. Alternatively, in decentralized approaches, no center controller is needed and all the functions are independently performed by each UAV or user. Compared to the centralized network, each UAV (user) needs to independently perceive the network state and spectrum state, and then make a choice by itself. This is particular challenging to be achieved since massive UAVs are connected via wireless links.

Game theory is a powerful method to realize distributed optimization, where each player with complex interactions spontaneously and independently makes its decision. Modeling the power control problem with game theory, especially non-cooperative games, has been widely investigated over the past few decades\cite{HZB}. However, these models should be re-evaluated in UAV-supported UDNs due to the dynamics of UAVs. Some dynamic games, e.g., repeated games, stochastic games, and differential games, might be useful in this case.
\section*{Case Study}
\subsection*{UAV-Supported BS}
\begin{figure}
\centering
\subfigure[]{
\label{BS_Model}
\includegraphics[width=0.4\textwidth]{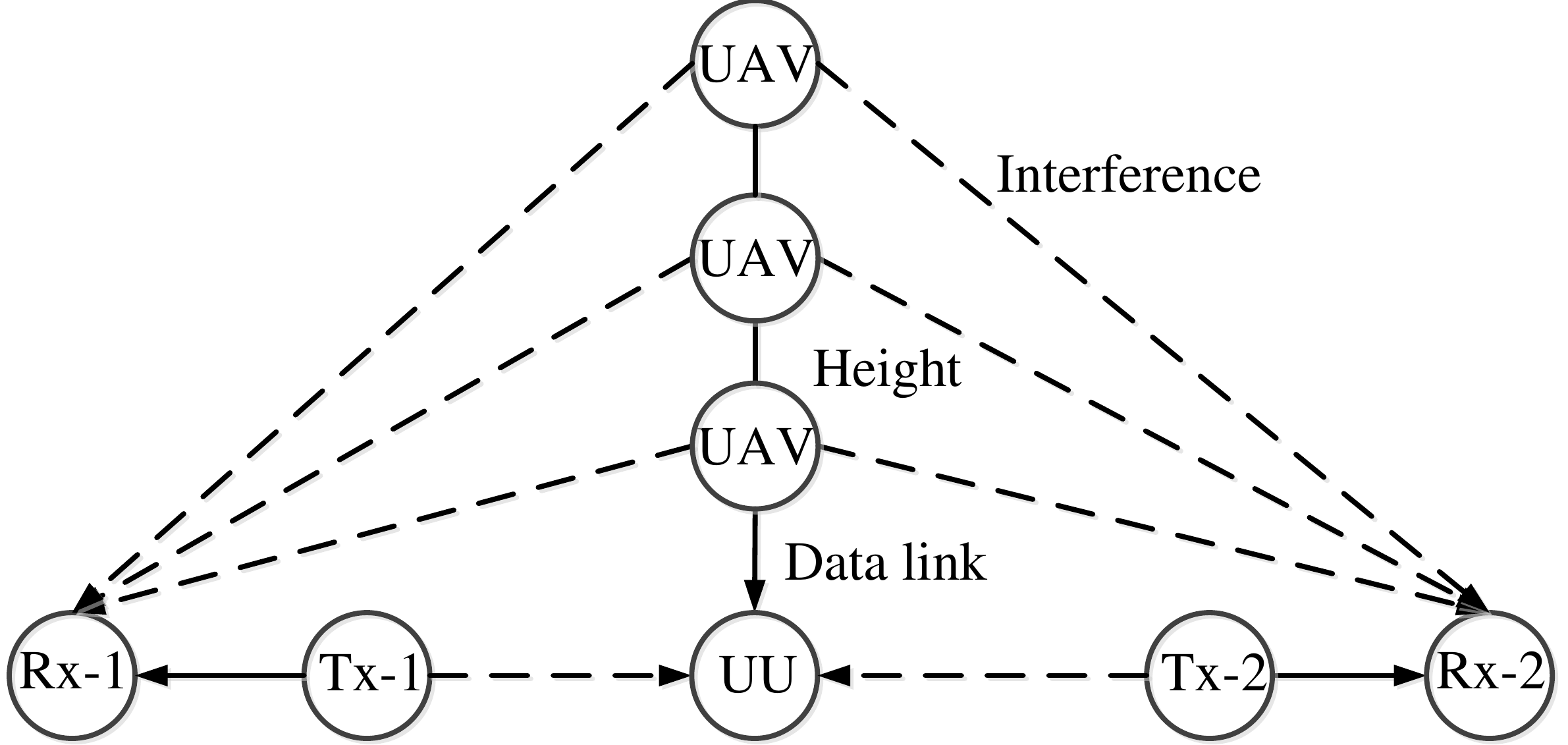}
}
\subfigure[]{
\label{BS_Simul}
\includegraphics[width=0.4\textwidth]{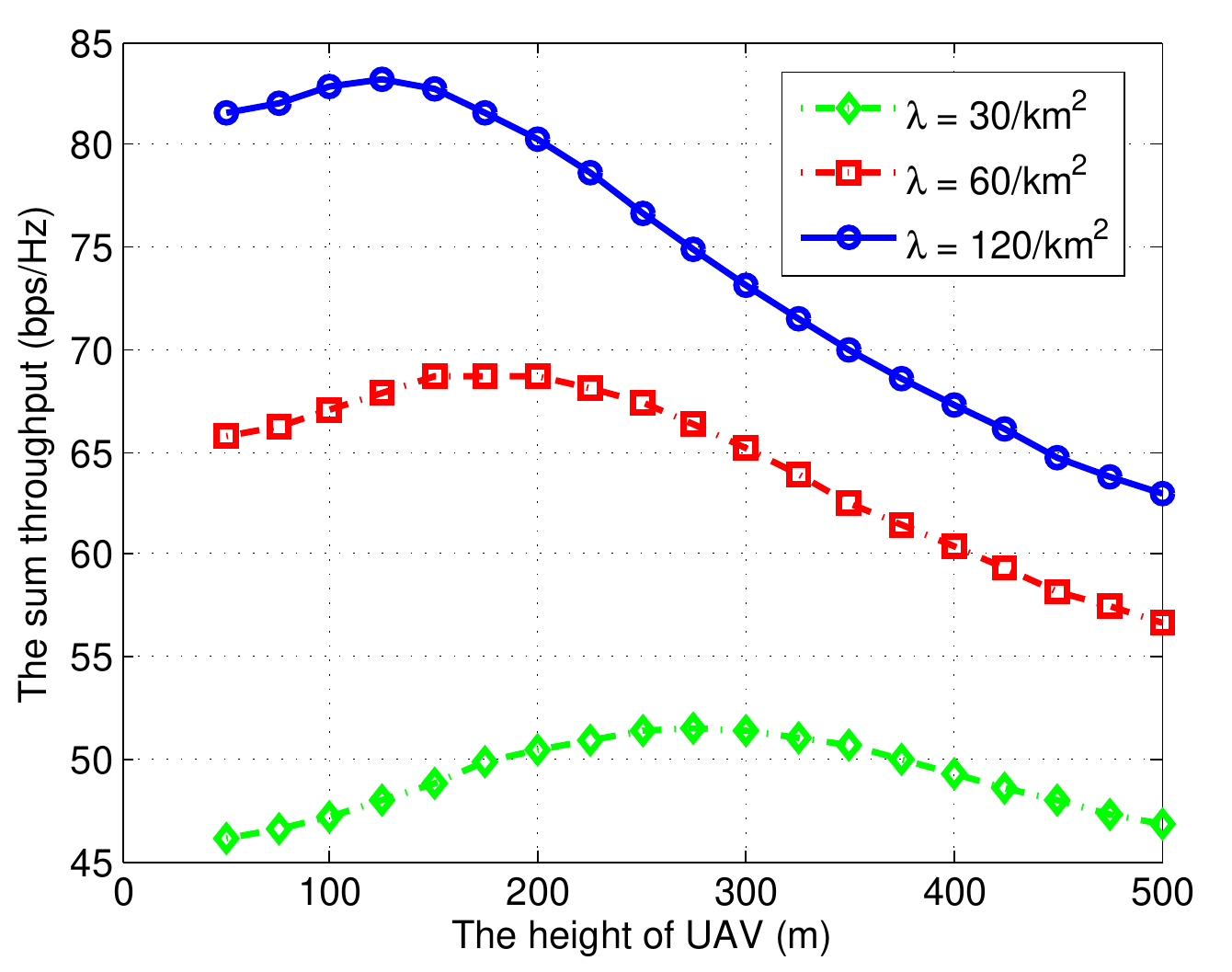}
}
 \caption{The UAV serves as a flying BS and the corresponding throughput performance: a) a considered scenario that a UAV serves as a flying BS; b) the sum throughput with different heights of UAV.}
 \label{fig:1}
\end{figure}

Consider a $1000$ m $\times$ $1000$ m scenario that a UAV hovering over the midpoint serves as a flying BS to provide wireless access for the UAV user, as shown in Fig. \ref{BS_Model}. Meanwhile, other communication links (D2D pairs) with the maximum distance as $D=30$ m share the same spectrum with the UAV user. The goal is to maximize the sum throughput while satisfying the signal to interference plus noise ratio requirement $\gamma$ of the UAV user, where $\gamma=5$ dB. The maximum transmit power of the UAV and D2D pairs are respectively 5 W and 100 mW. For the D2D transmission, we consider a Rayleigh fading channel model as in\cite{MM2}. The channel power gain from the UAV to the ground user is considered to be dominated by LOS as in\cite{relaytc}. By optimizing the transmit power of each link and the height of UAV to mitigate the mutual interference, the system throughput can be greatly improved. Figure. \ref{BS_Simul} illustrates the system throughput with different heights of UAV under various user densities $\lambda$. It can be observed that the system throughput grows with the user density. Moreover, deploying the UAV as flying BS can bring additional performance improvement by adjusting its height. Specifically, strong information signal between the UAV and UAV user generally means serious interference induced by UAV, and vice versa. Therefore, there is an optimal point of the height that balances the received information signal and induced interference. Moreover, the optimal height gradually decreases with increasing user density. This is due to the fact that the interference experienced by the UAV user becomes serious with increasing user density and thus the UAV must reduce its height to provide better channel state.
\subsection*{UAV-Supported Relay}
\begin{figure}
\centering
\subfigure[]{
\label{Ralay_Model}
\includegraphics[width=0.6\textwidth]{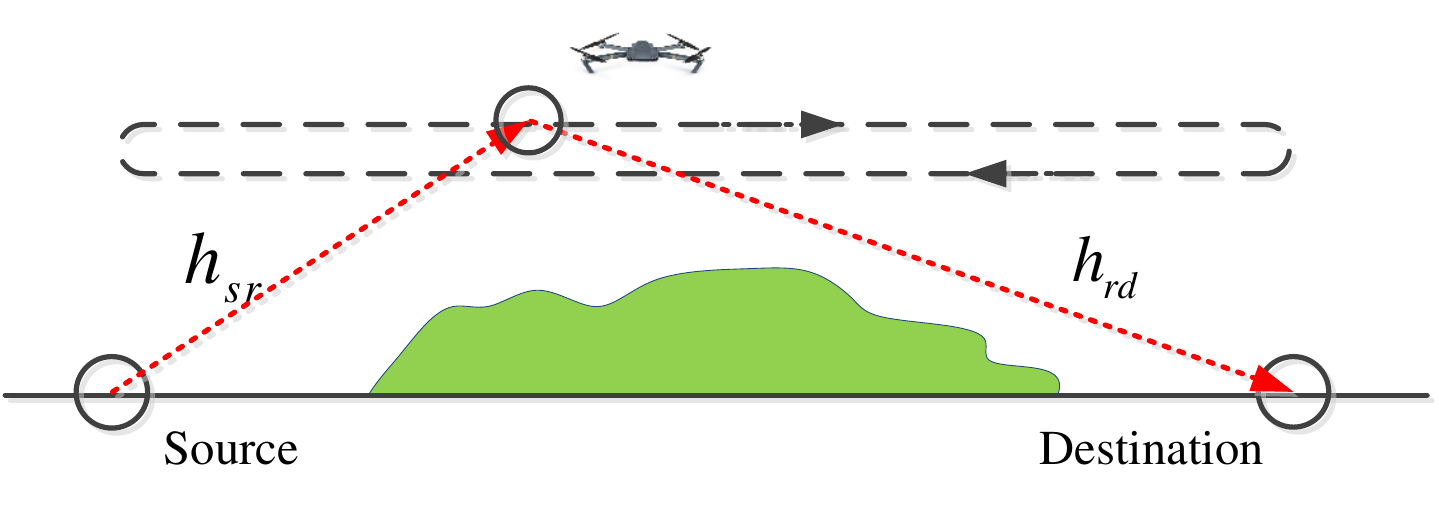}
}
\subfigure[]{
\label{Relay_Simul}
\includegraphics[width=0.6\textwidth]{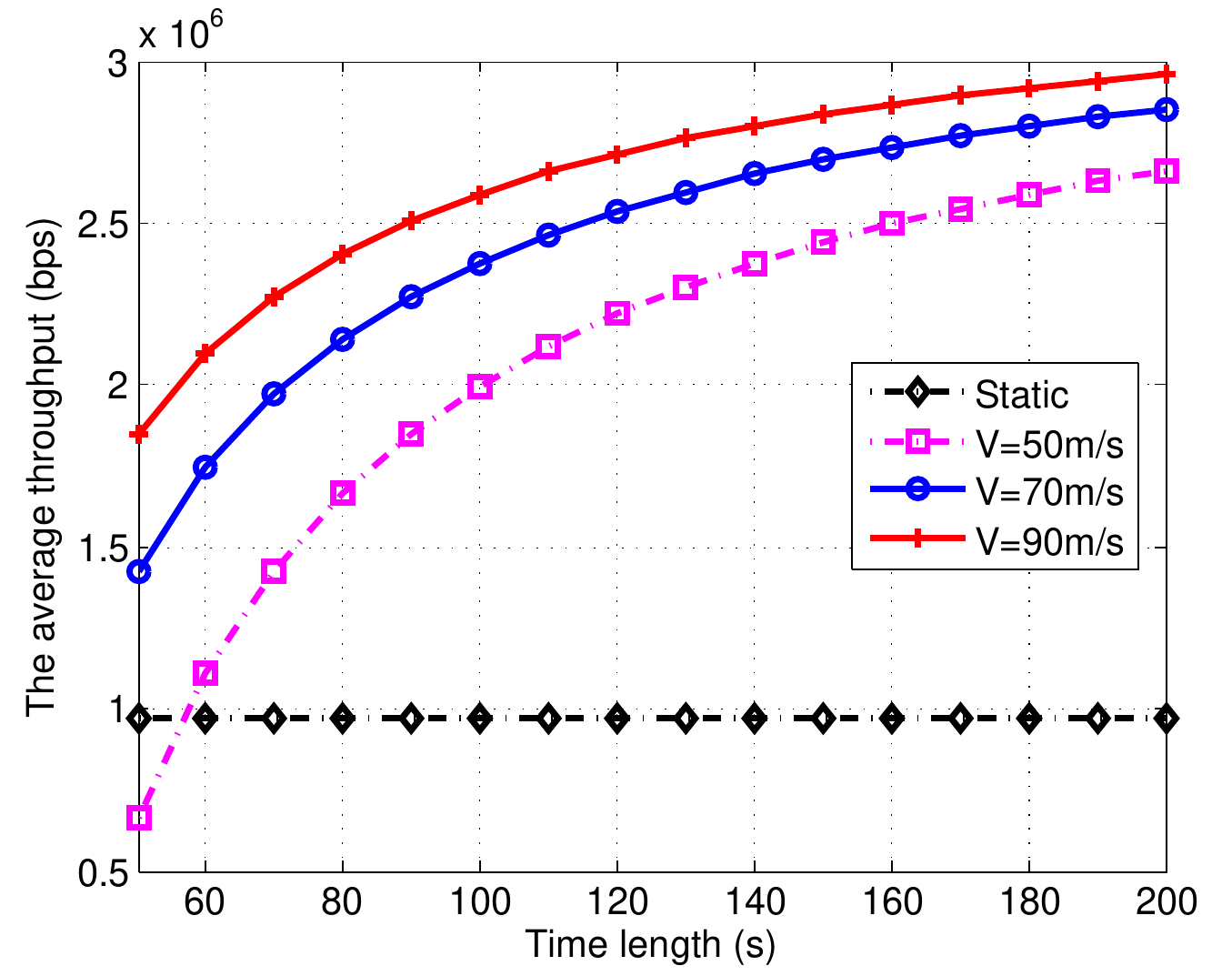}
}
\subfigure[]{
\label{Relay-EE}
\includegraphics[width=0.6\textwidth]{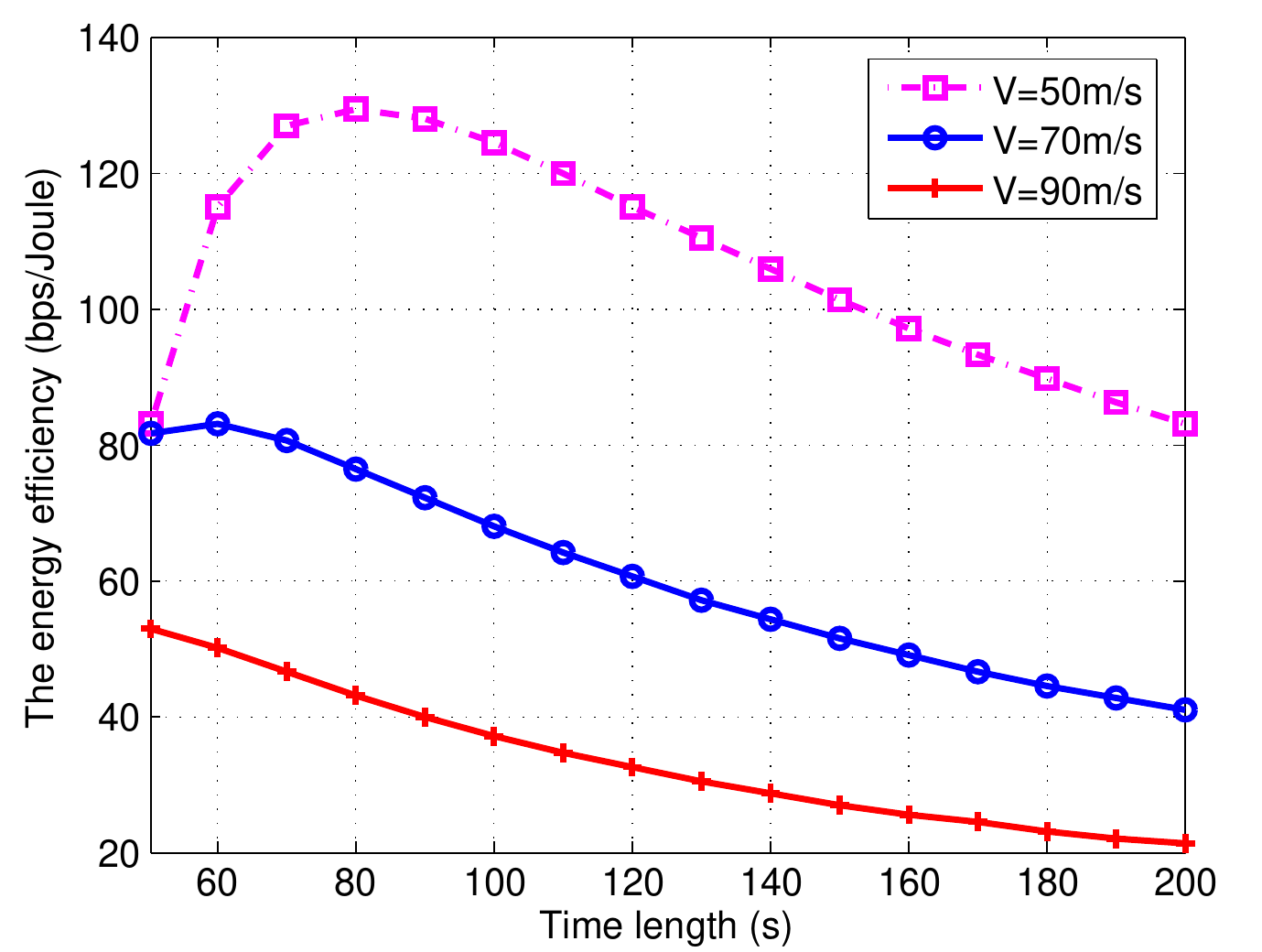}
}
\caption{The UAV serves as mobile relay and the corresponding system performance: a) a considered scenario that a UAV serves as a mobile relay; b) the average throughput with different time lengths; c) the energy efficiency performance.}
\label{fig:1}
\end{figure}

Figure. \ref{Ralay_Model} illustrates a considered mobile relay system with the distance between a source node and a destination node as $L=2000$ m, where the quality of direct link cannot meet the communication requirements due to e.g., limited backhaul or severe obstacle. The nodes can be ground stations (smallcells or macrocells) without ideal backhauls or users far from apart. Thus, a UAV of sufficiently high mobility is employed to assist the communication link within a certain time length $T$. During the flight, the UAV flying at a fixed altitude $H=100$ m can receive the information from the source node, and at the same time, it forwards the received information to the destination node. The transmit power and trajectory are jointly optimized to exploit the channel variations. Performance improvement can be achieved by deploying a mobile relay as shown in Fig. \ref{Relay_Simul}, where a static relay as the benchmark is located at the midpoint. It can be observed that the average throughput grows with flight velocity $V$ and/or time length $T$. The reason can be illustrated as follows: Under the situation of larger $V$ and/or longer $T$, there are more time for the UAV to hover above the nodes, which means best channel gain can be achieved, thereby realizing better performance.

To further evaluate the performance of mobile relay systems, the energy efficiency is plotted in Fig. \ref{Relay-EE}, where an energy consumption model with straight and level flight is considered as in\cite{EE}. Since the mechanical energy consumption is the dominant part of the UAV energy consumption, the communication related energy consumption is ignored. More accurate energy consumption model will be left as future work. It can be observed that the energy efficiency decreases with increasing velocity while the throughput grows. This is due to the fact that high velocity means much more energy consumption and limited throughput gain. Therefore, there is a tradeoff between throughput gain and energy efficiency, which means that the UAV's trajectory and transmit power should be jointly studied.
\subsection*{UAV-Supported Energy Transfer}
\begin{figure}[t]
\begin{center}
\subfigure[]{\includegraphics[width=0.45\textwidth]{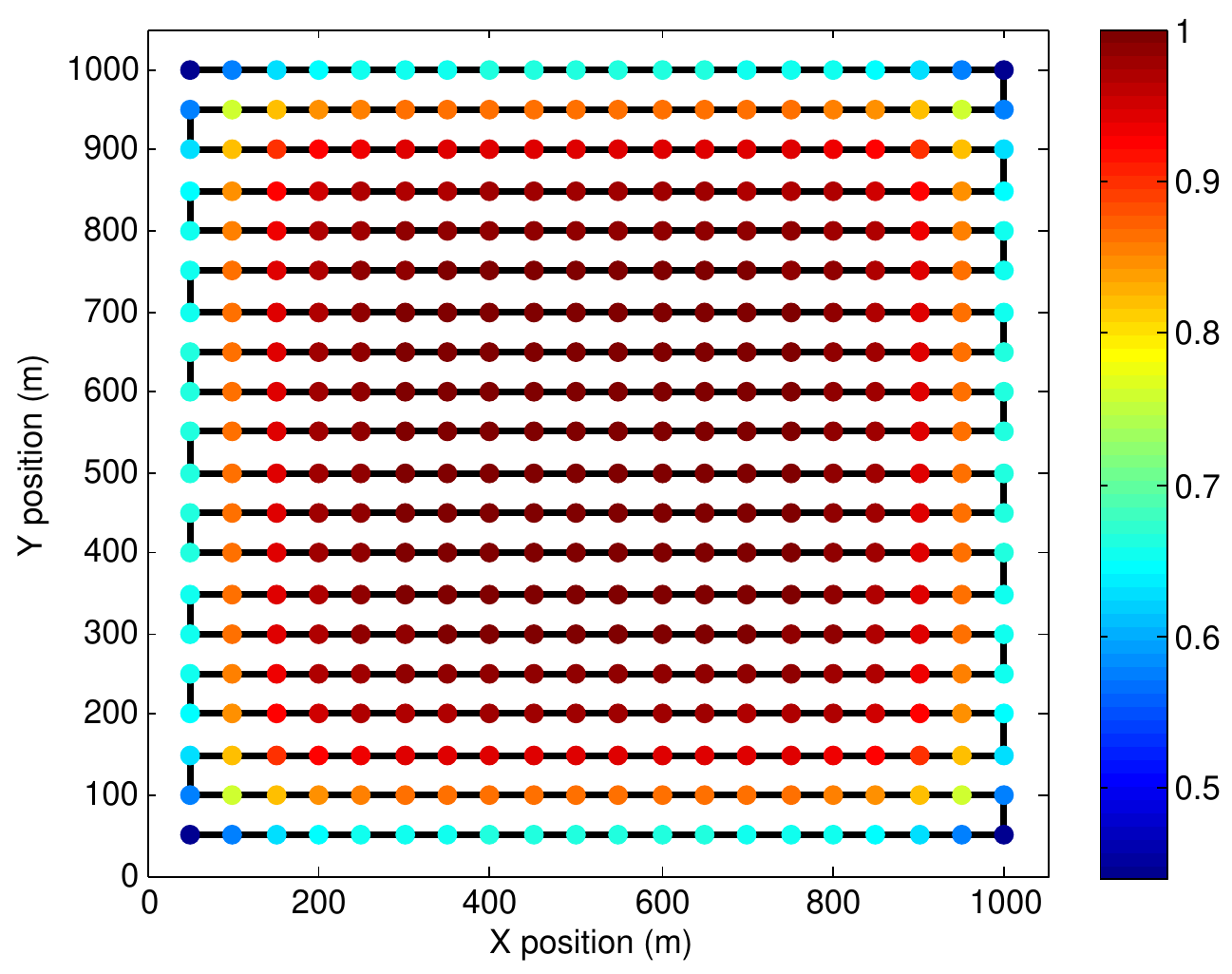}}
\subfigure[]{\includegraphics[width=0.45\textwidth]{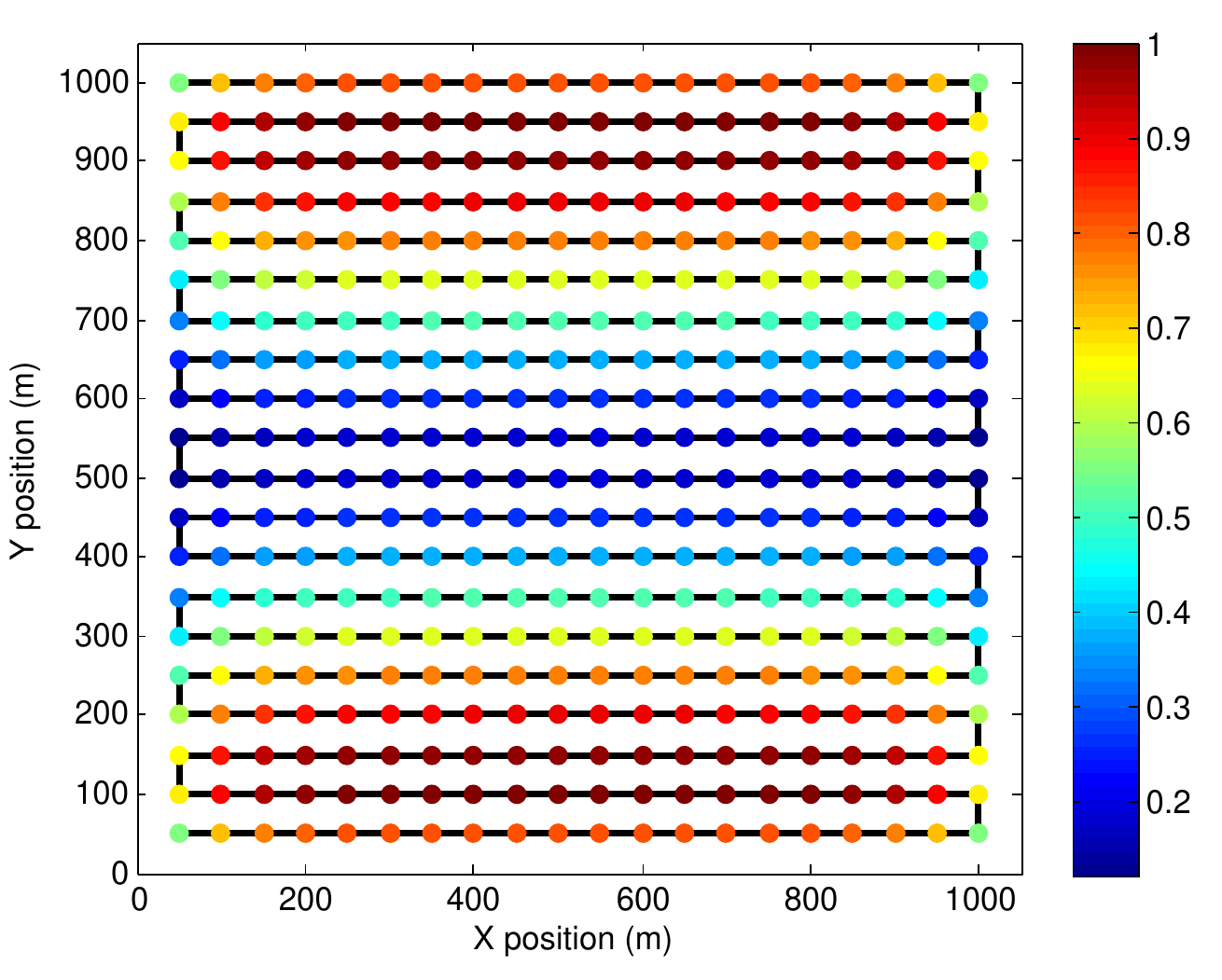}}
\\
\subfigure[]{\includegraphics[width=0.45\textwidth]{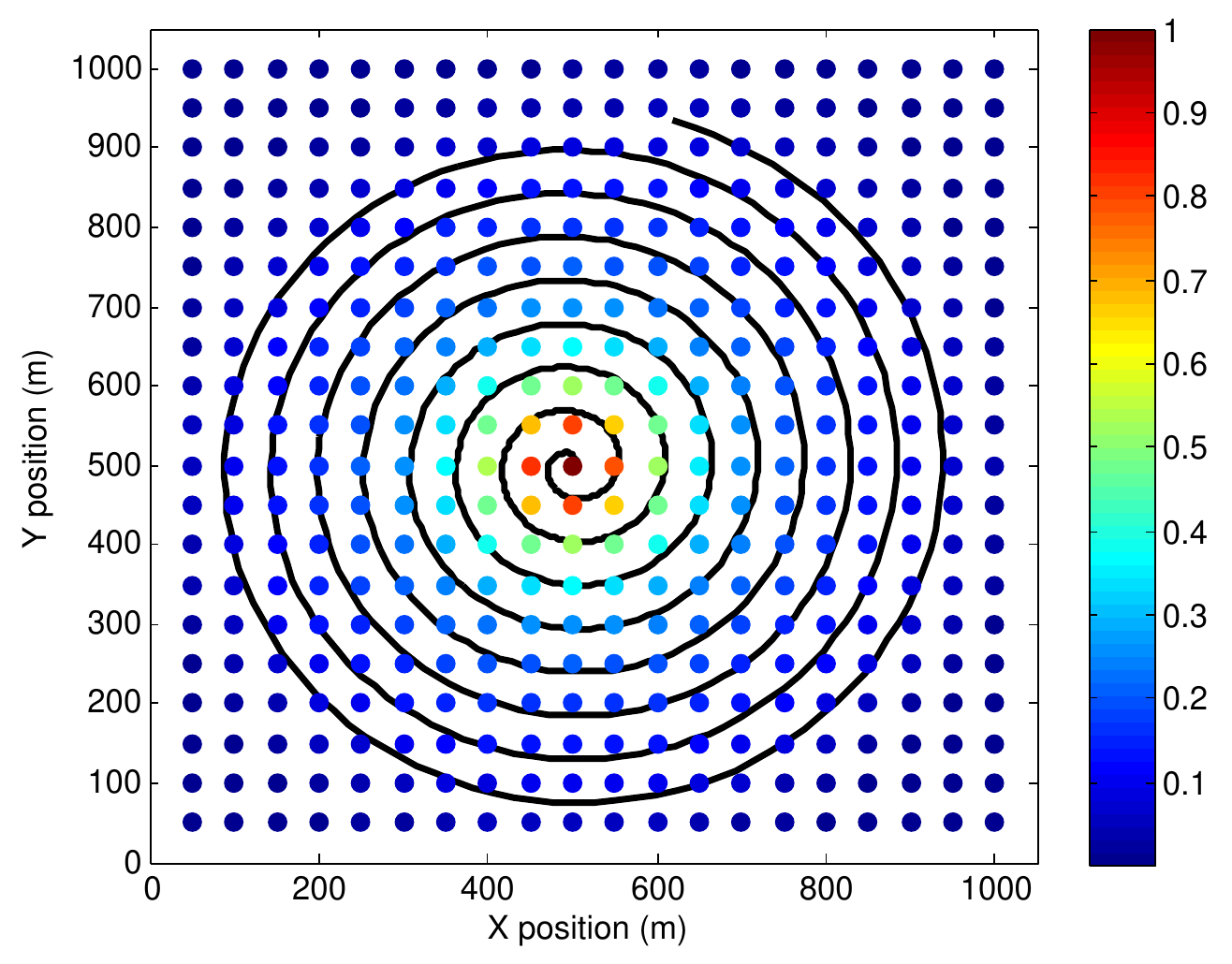}}
\subfigure[]{\includegraphics[width=0.45\textwidth]{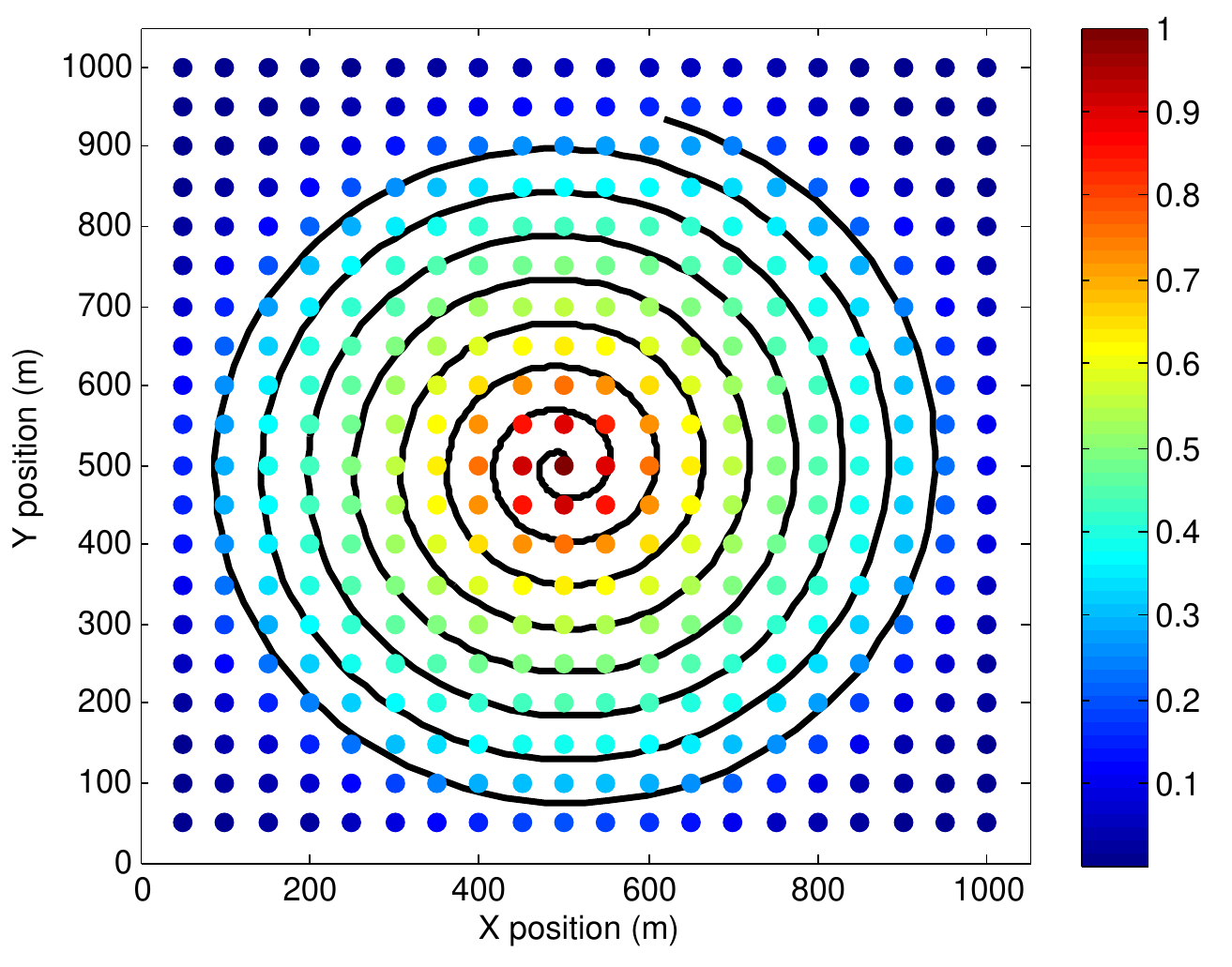}}
\caption{Normalized harvested energy under different trajectories with power control: a) sigmoid trajectory with fixed power; b) sigmoid trajectory with time varying power; c) spiral trajectory with fixed power; d) spiral trajectory with time varying power.}
\label{EH_result}
\end{center}
\end{figure}

As illustrated in Fig. \ref{EH_result}, $20 \times 20$ nodes are uniformly located in a $1000$ m $\times$ $1000$ m area, where a UAV flying at a fixed altitude $H=100$ m provides wireless energy for ground nodes. These low-power nodes can be viewed as sensor nodes, which possess much important information to be transmitted, however, it is considered that these nodes do not have any fixed energy source. Further, for satisfying the energy requirements of the nodes, the energy harvesting technology is employed. The normalized harvested energy under different trajectories (sigmoid and spiral trajectories with solid lines) with power control is presented  in Fig. \ref{EH_result}, where the power is fixed in (a) and (c), decreases first and then increases in (b), and always increases in (d). With a flying energy source and power control, the harvested energy for each node can be easily controlled. Given the transmit power, adjusting the UAV's trajectory results in the diversity of harvested energy, such as (a) and (c). Furthermore, given the UAV's trajectory, one can fine-tune the power to accomplish users' requirements, e.g., (a) and (b) or (c) and (d). It is quite interesting in the future work to further consider that the energy consumption rate, remaining energy, and the required energy of each node in UDNs are distinguishing.
\subsection*{UAV-Supported Caching}
\begin{figure}
\centering{\includegraphics[width=120mm]{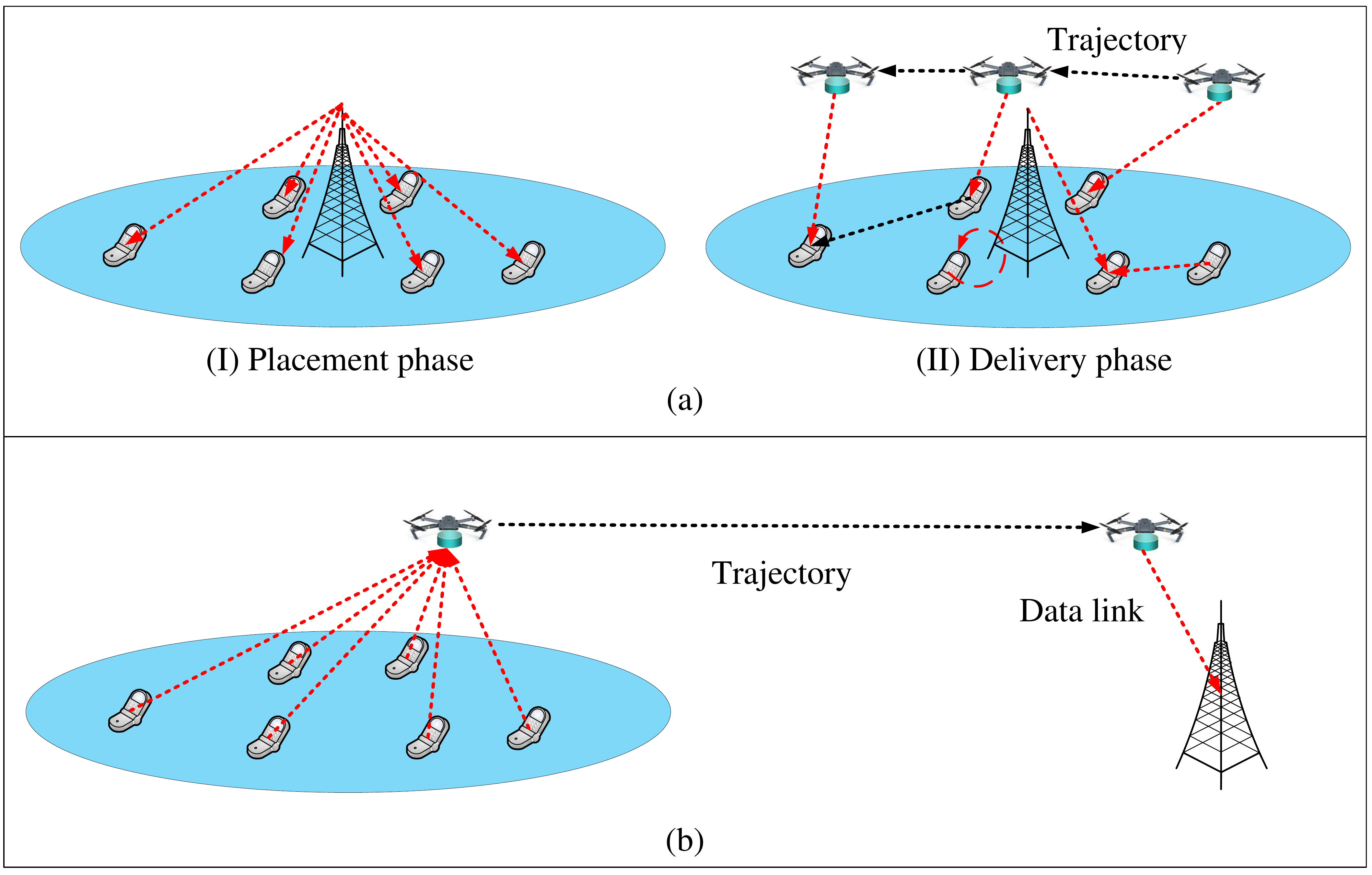}}
\caption{The UAV serves as mobile cache under two scenarios: a) serving mobile users; b) collecting and transmitting the information.}
\label{Caching_M}
\end{figure}

Figure. \ref{Caching_M} shows two scenarios that a UAV serves as a mobile cache. The two-phase protocol for serving mobile users is illustrated in Fig. 6(a). In the first phase (placement phase), each user independently and randomly caches one out of $N$ contents in its memory from the BS or other information source. Then, each user requests one content in the delivery phase. If a user can find its requested content in its own memory, the user acquires the content instantaneously without any transmission. However, once the requested content cannot be found in its own memory, it must send a request to the UAV or other users through a D2D transmission. Comparing to static caching, the flying UAV can provide better quality of experience with tracking the mobility of users. Since the UAV can move with the users, the distance between them reduces, which means that it achieves better CSI and reduces the delay. Moreover, the mobile users can be always served by one UAV, which avoids frequently switching among multi-cells.

Figure. 6(b) illustrates the application of the UAV as a mobile cache in delay-tolerant surveillance. The UAV first collects the information from the ground sensors and caches these data. Then, the UAV flies to the ground center and transmits the information back to the center.
\section*{Challenges and Open Issues}
To achieve the potentials of UAV-supported UDNs, there are some roadblocks that urgently need to be addressed.
\begin{itemize}
  \item \textbf{Highly time-varying network topology and channel state:} In contrast to the terrestrial networks, the high mobility of UAVs results in frequently changing network topology. Moreover, in addition to the vibration resulting from small scale fading, the large scale fading of the channel state changes greatly. Acquiring accurate CSI in real time is challenging.
  \item \textbf{Coordination among massive UAVs:} The coordination must be considered for effective UAVs deployment. First, accident-free operation is a prerequisite for performing any tasks. To this end, UAVs require precise coordination to avoid collision or address the blockage. Moreover, they should cooperate to improve network performance, such as reducing interference.
  \item \textbf{Limited energy of the UAVs:} The limited energy is a major challenge since UAVs flying in the air have no fixed energy supply. For one thing, the mechanical power consumption is the dominant part of the UAV power consumption and limits significantly the applications. For another, the communications among the UAVs or between the UAV and central controller also consume much energy, which is even higher than that is used to serve users.

\end{itemize}

These challenges motive us to carefully examine the power control issues in UAV-supported UDNs. Several open issues are presented as follows:
\begin{itemize}
  \item \textbf{Flying ad hoc networks (FANETs):} The communications among massive UAVs require a self-organized network. Although mobile and vehicular ad hoc networks have been widely investigated, these works can not answer the questions in FANETs, such as high mobility and limited energy.
  \item \textbf{Efficient deployment:} Although the cost of UAVs is declining, deploying large number of UAVs are still costly. The efficient deployment is important but challenging  since massive various UAVs are connected and interfere. To serve the users more efficiently, the number, locations, stored energy of UAVs should be considered to assist the power control strategy.
  \item \textbf{Interference management:} Considering UDNs that massive users operate in co-channel scenario, serious interference undoubtedly drives out the densification gains. As the topology of UAV-supported UDN changes rapidly, interference will be more complex in a dynamic environment. Low complexity and high performance interference management methods are expected.
\end{itemize}
\section*{Conclusion}
In this article we present an overview of power control in UAV-supported UDNs. Four representative scenarios, i.e., aerial BS, mobile relay, energy transfer, and caching, are investigated with simulations, where it has been shown that additional performance gain can be achieved by introducing UAVs. We then focus on the efficient power control from the perspectives of design considerations and methods. Furthermore, we particularize the challenges and discuss several future research directions. We firmly believe this area will be a fruitful research direction and we have just touched one tip of the iceberg. We hope this article will stimulate much more research interests.

\section*{Acknowledgments}
This work is supported by the National Natural Science Foundation of China (Grant No. 61501510), Natural Science
Foundation of Jiangsu Province (Grant No. BK20150717), China Postdoctoral Science Funded Project (Grant No. 2016M590398),
and Jiangsu Planned Projects for Postdoctoral Research Funds (Grant No. 1501009A).

\section*{Biographies}
Haichao Wang (whcwl0919@sina.com) received the B.S. degree in electronic engineering from the College of Communications Engineering, Nanjing, China, in 2014. He is currently pursuing the Ph.D. degree in communications and information system in College of Communications Engineering. His research interests focus on interference mitigation techniques, green communications, UAV communications, and convex optimization techniques.

Guoru Ding (dr.guoru.ding@ieee.org) is an assistant professor in College of Communications Engineering and a Postdoctoral Research Associate at the National Mobile Communications Research Laboratory, Southeast University, Nanjing, China. He received his B.S. degree from Xidian University in 2008 and his PhD degree from the College of Communications Engineering, Nanjing, China, in 2014. His research interests include cognitive radio networks, massive MIMO, machine learning, and big data analytics over wireless networks.

Feifei Gao (feifeigao@ieee.org) received the Ph.D. degree from National University of Singapore, Singapore, in 2007. He was a Research Fellow with the Institute for Infocomm Research, A*STAR, Singapore, in 2008 and was an Assistant Professor with the School of Engineering and Science, Jacobs University, Bremen, Germany, from 2009 to 2010. In 2011, he joined the Department of Automation, Tsinghua University, Beijing, China, where he is currently an Associate Professor.

Jin Chen (chenjin99@263.net) received the B.S. degree in communications engineering and M.S. and Ph.D. degrees in communications and information system from the Institute of Communications Engineering, Nanjing, China, in 1993, 1996, and 1999, respectively. She is currently a Professor with the PLA Army Engineering University, Nanjing. Her research interests include cognitive radio networks, distributed optimization algorithms, and digital signal processing.

Jinlong Wang (wjl543@sina.com) received the B.S. degree in wireless communications and M.S. and Ph.D. degrees in communications and electronic systems from the Institute of Communications Engineering, Nanjing, China, in 1983, 1986, and 1992, respectively. He is currently a Professor with the PLA Army Engineering University, Nanjing. His research interests span a wide range of topics in wireless communications and signal processing, including cognitive radio networks, HF communications, cooperative communications, and wireless security.

Le Wang (wlwhc0919@sina.com) received the B.S. degree in electronic engineering from College of Communications Engineering, Nanjing, China, in 2014. She is currently pursuing the M.S. degree in communications and information system in College of Communications Engineering. Her research interests focus on resource allocation, channel estimate, and convex optimization techniques.

\newpage
\begin{figure}[!t]
\centering{\includegraphics[width=140mm]{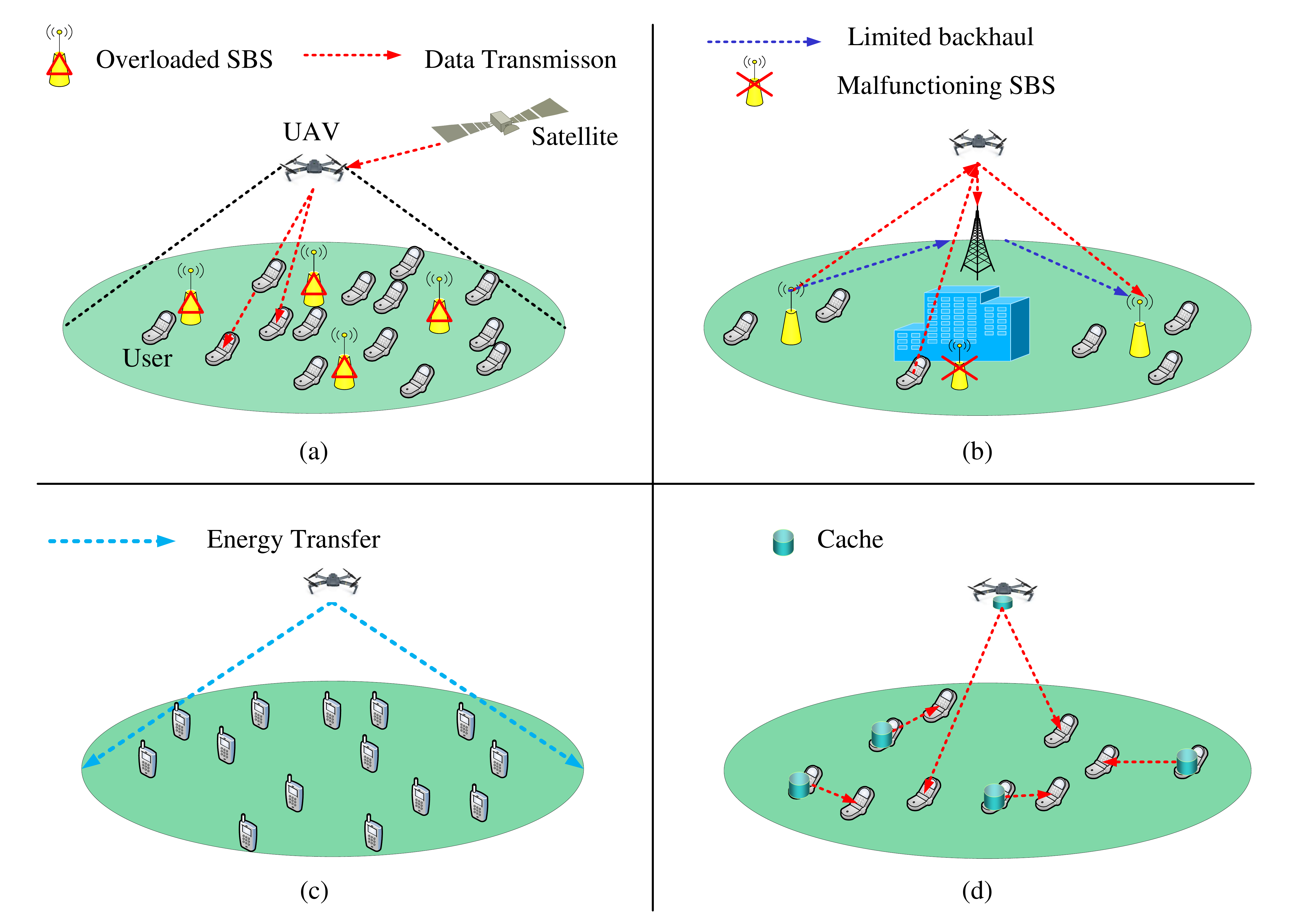}}
\caption{Four representative cases for UAV-supported UDNs: a) UAV-supported BS; b) UAV-supported relay; c) UAV-supported energy transfer; d) UAV-supported caching.}
\end{figure}
~\\
\newpage
\begin{figure}[!t]
\centering{\includegraphics[width=120mm]{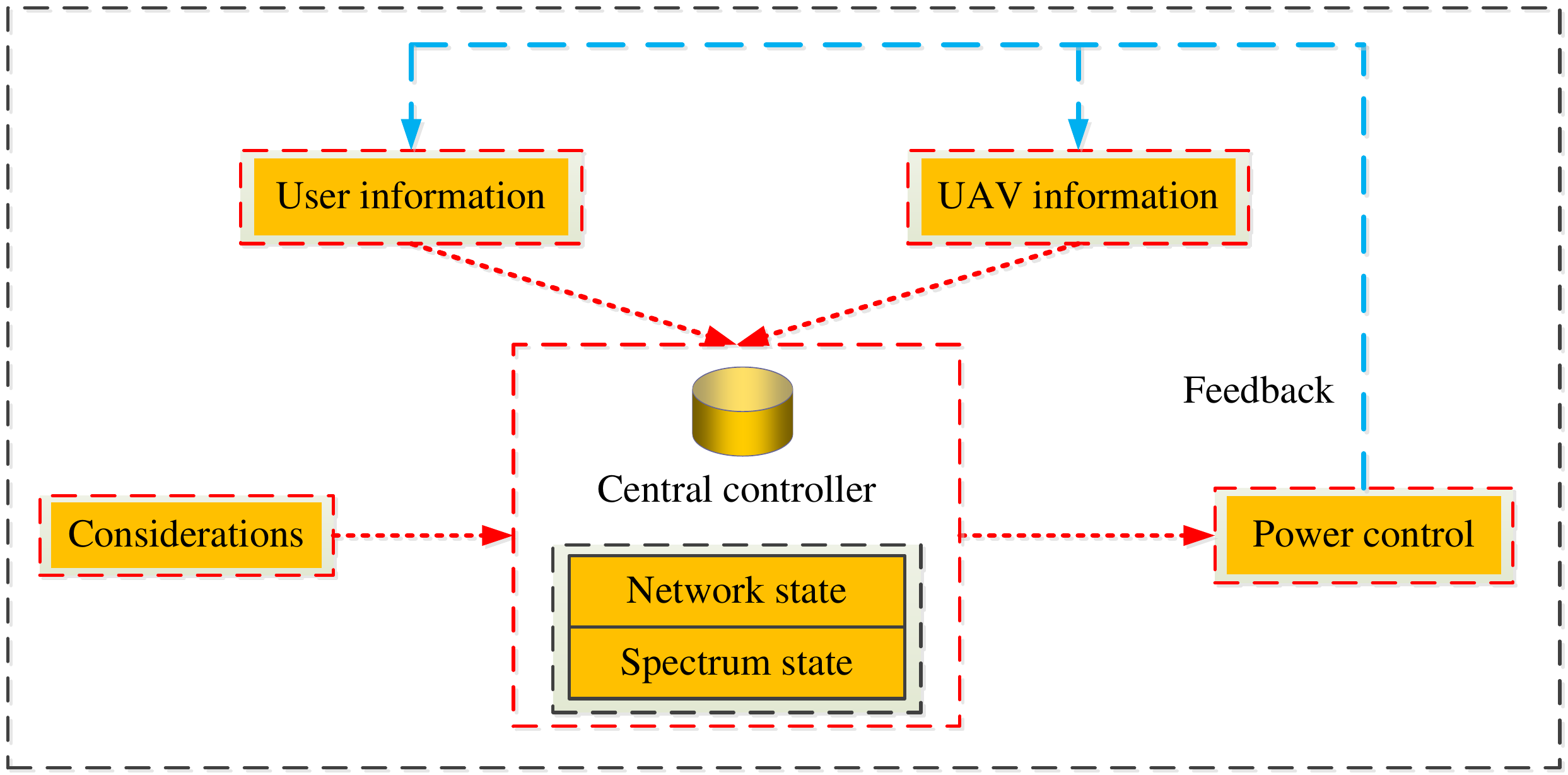}}
\caption{The centralized power control architecture for UAV-supported UDNs.}
\end{figure}
~\\
\newpage
\begin{figure}
\centering
\subfigure[]{
\includegraphics[width=0.4\textwidth]{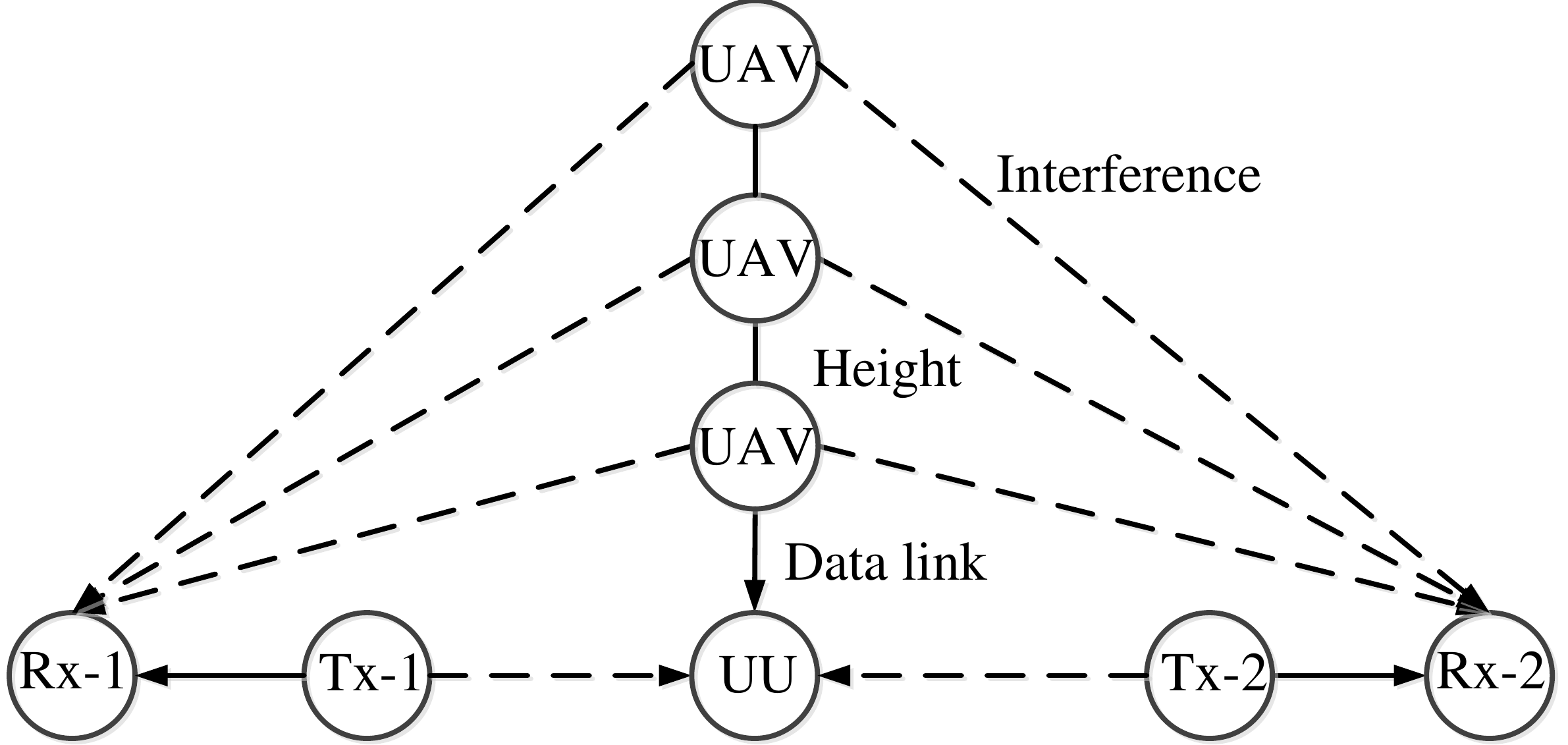}
}
\subfigure[]{
\includegraphics[width=0.4\textwidth]{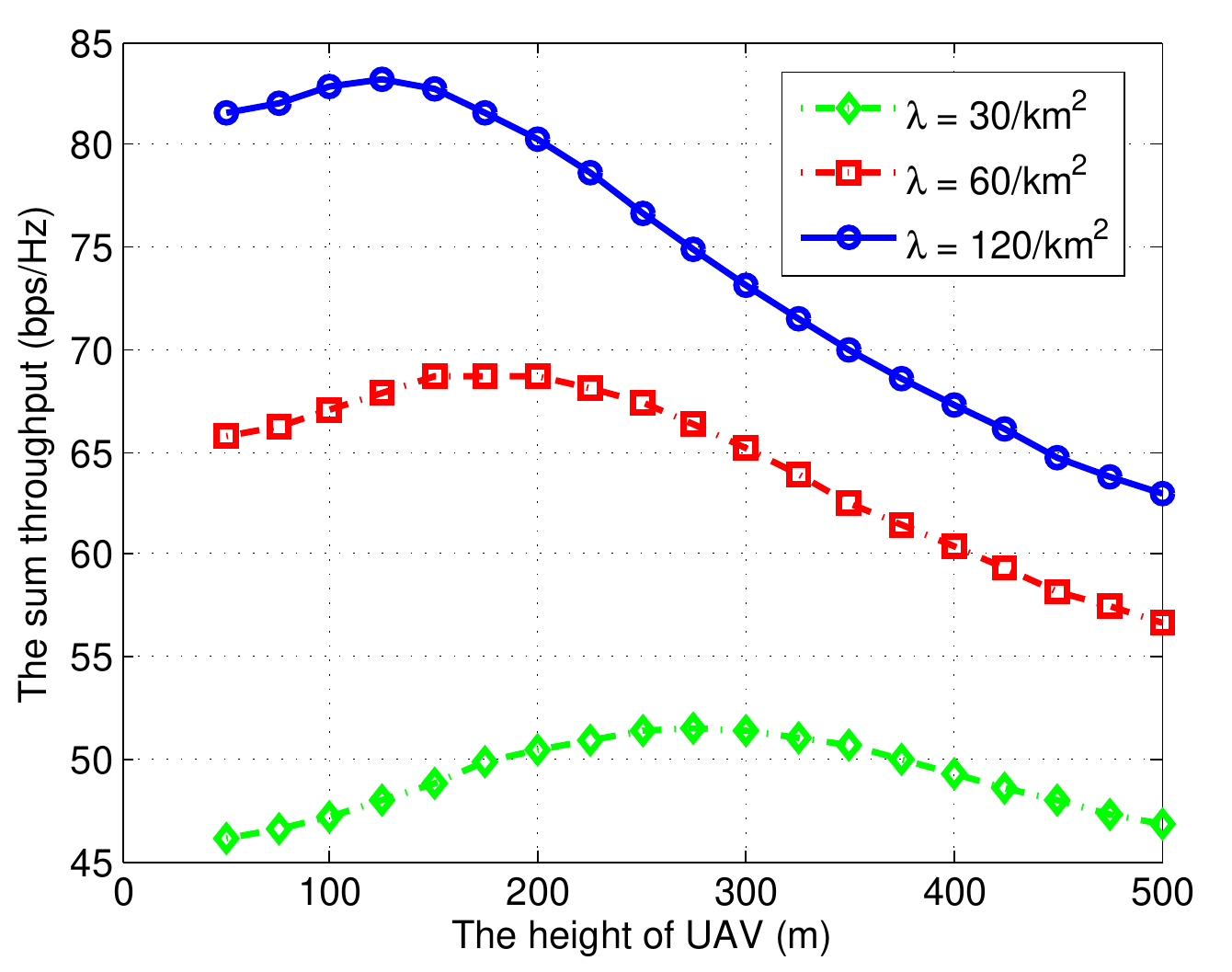}
}
 \caption{The UAV serves as a flying BS and the corresponding throughput performance: a) a considered scenario that a UAV serves as a flying BS; b) the sum throughput with different heights of UAV.}
\end{figure}

\newpage
\begin{figure}
\centering
\subfigure[]{
\includegraphics[width=0.6\textwidth]{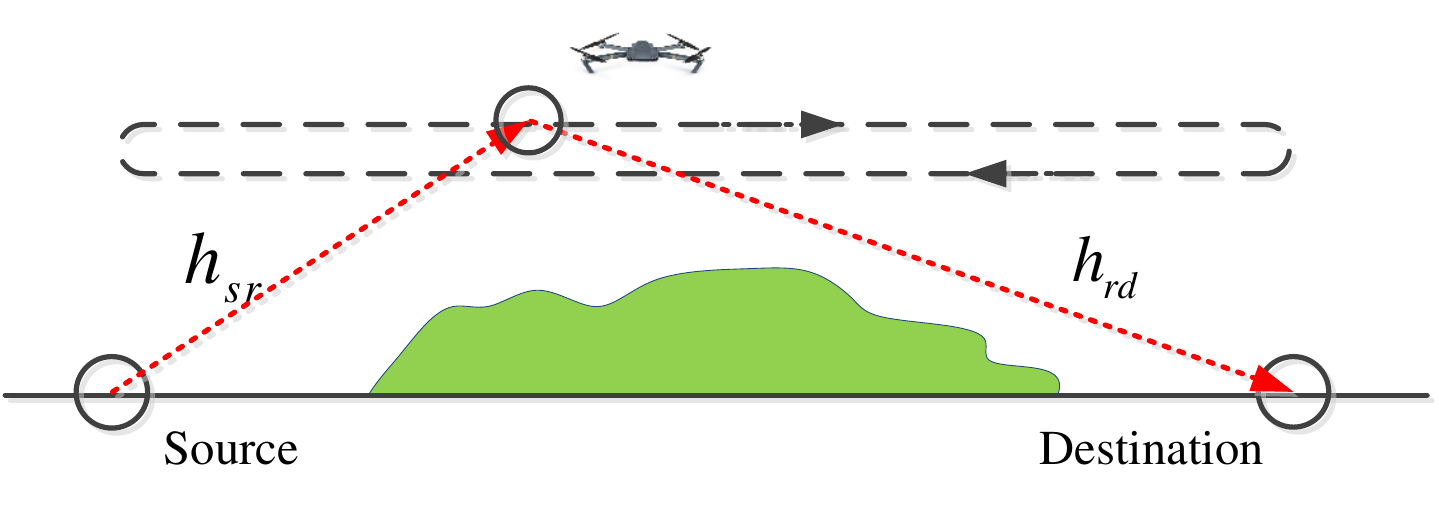}
}
\subfigure[]{
\includegraphics[width=0.6\textwidth]{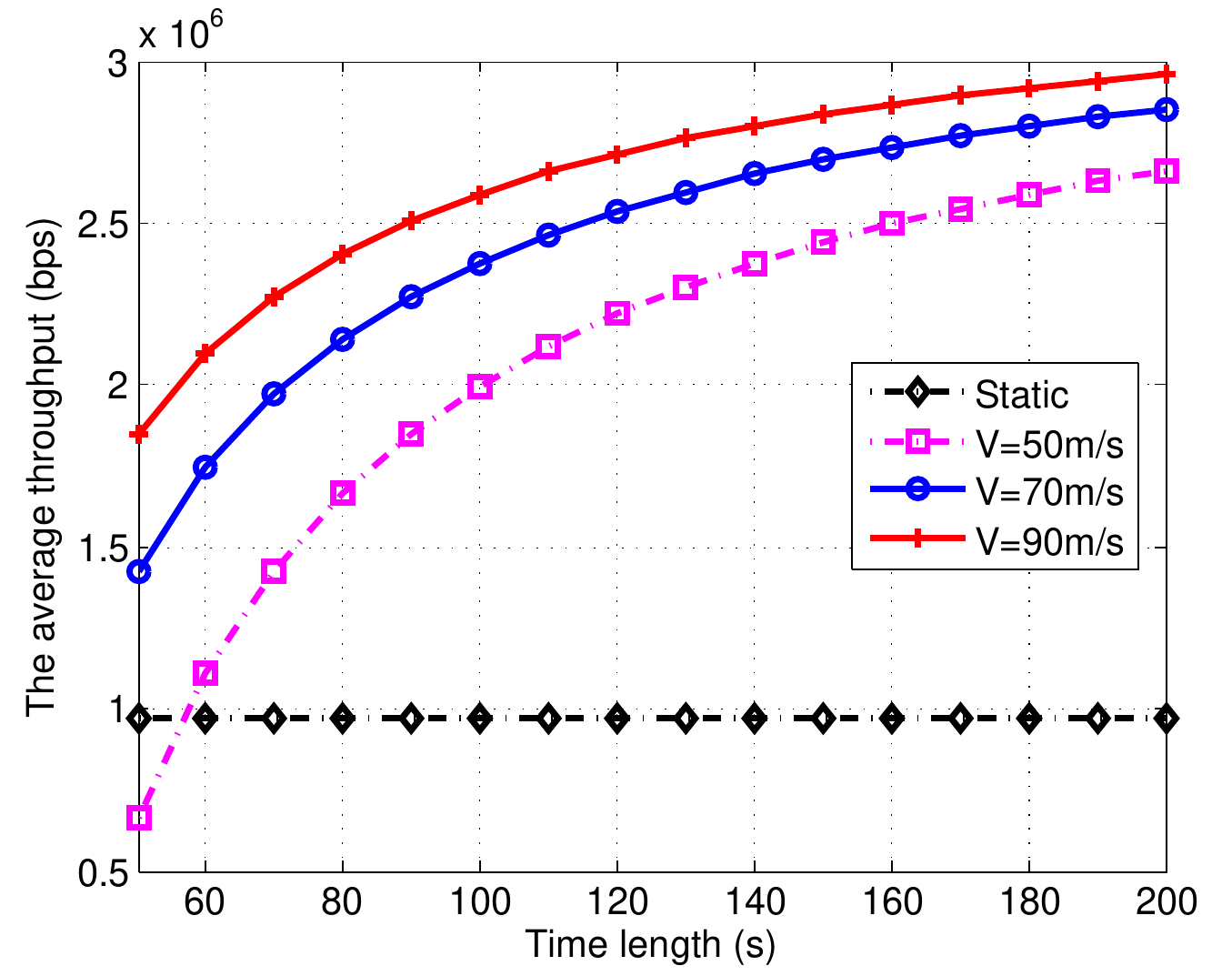}
}
\subfigure[]{
\includegraphics[width=0.6\textwidth]{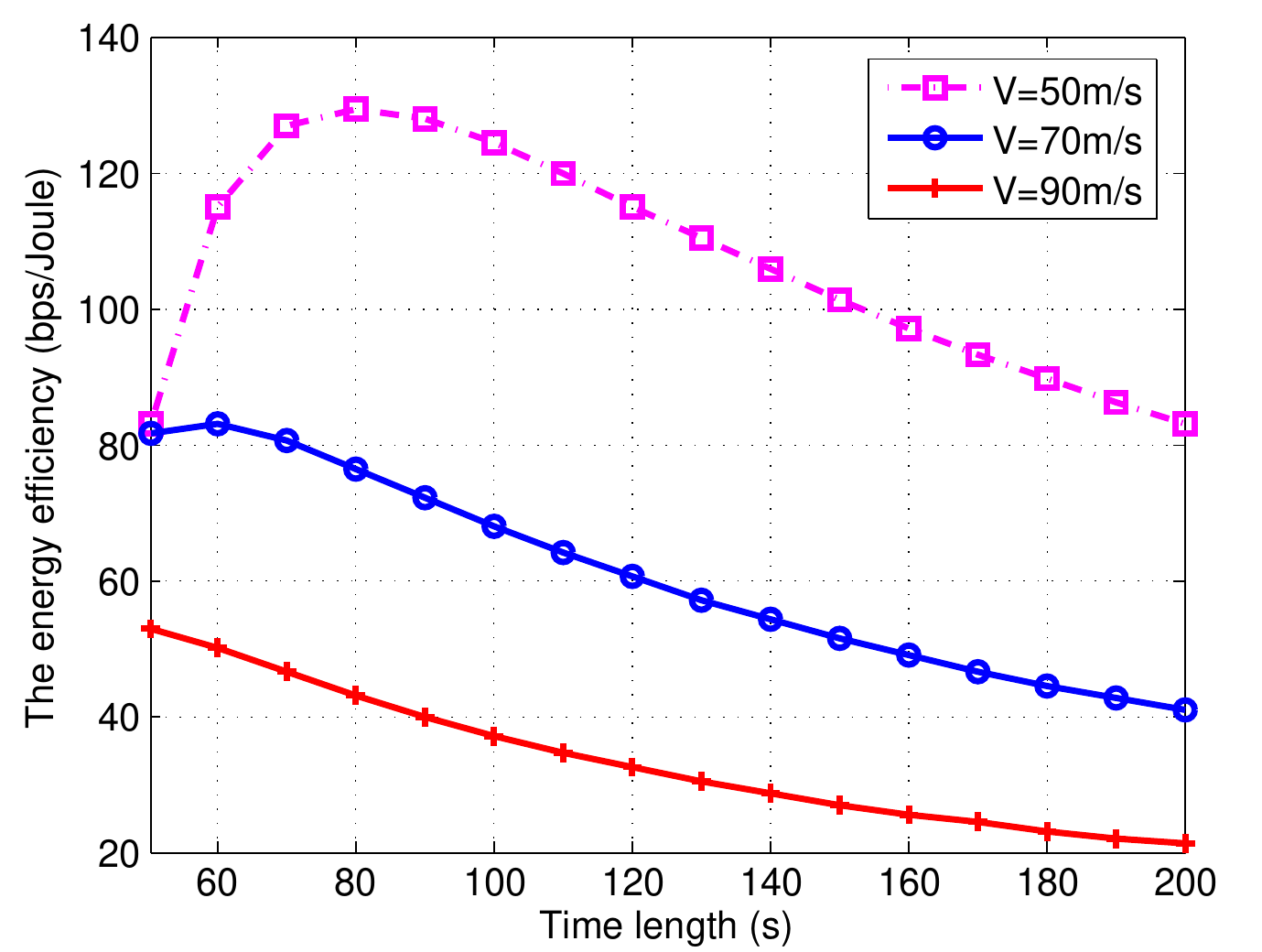}
}
\caption{The UAV serves as mobile relay and the corresponding system performance: a) a considered scenario that a UAV serves as a mobile relay; b) the average throughput with different time lengths; c) the energy efficiency performance.}
\end{figure}

\newpage
\begin{figure}[t]
\begin{center}
\subfigure[]{\includegraphics[width=0.45\textwidth]{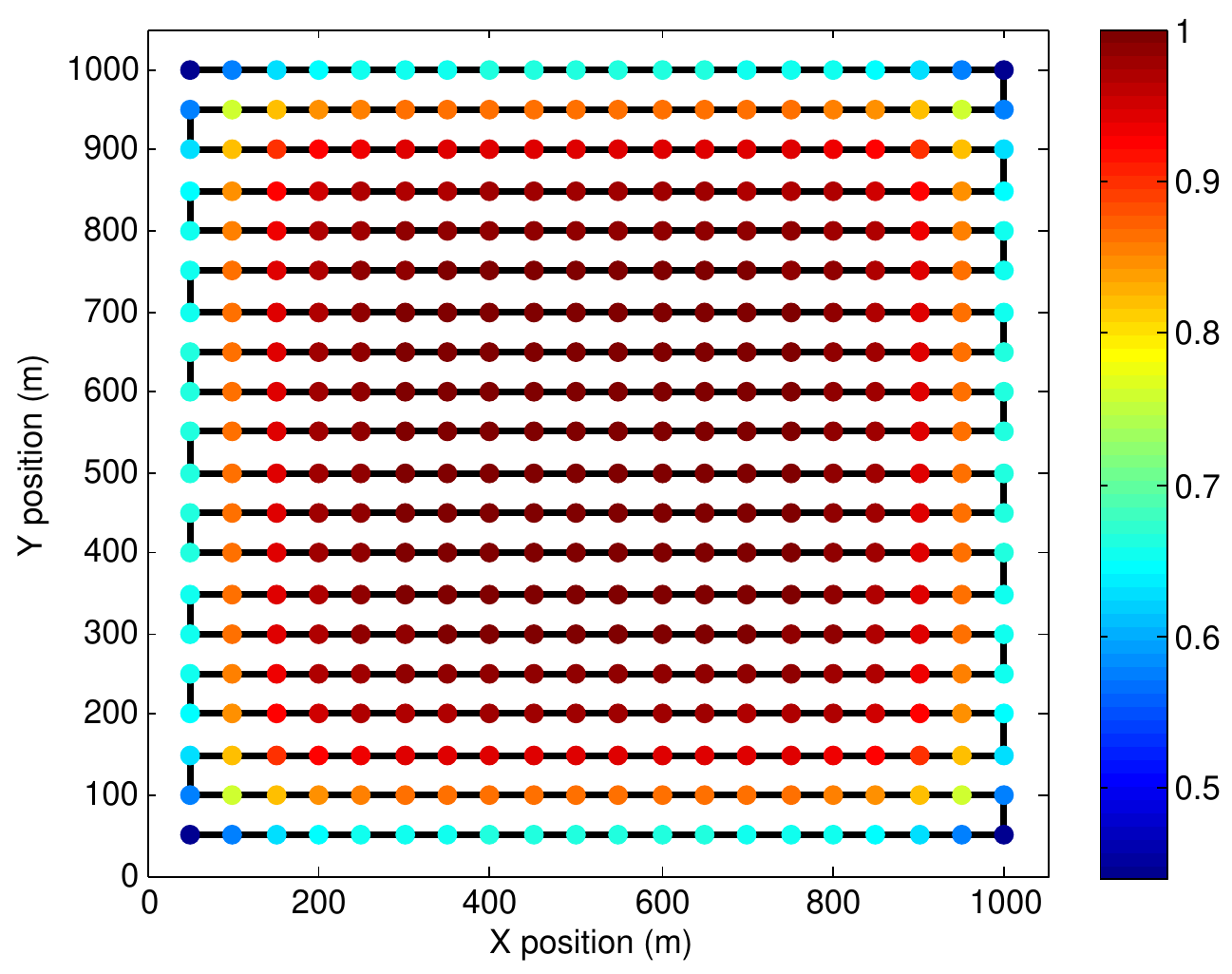}}
\subfigure[]{\includegraphics[width=0.45\textwidth]{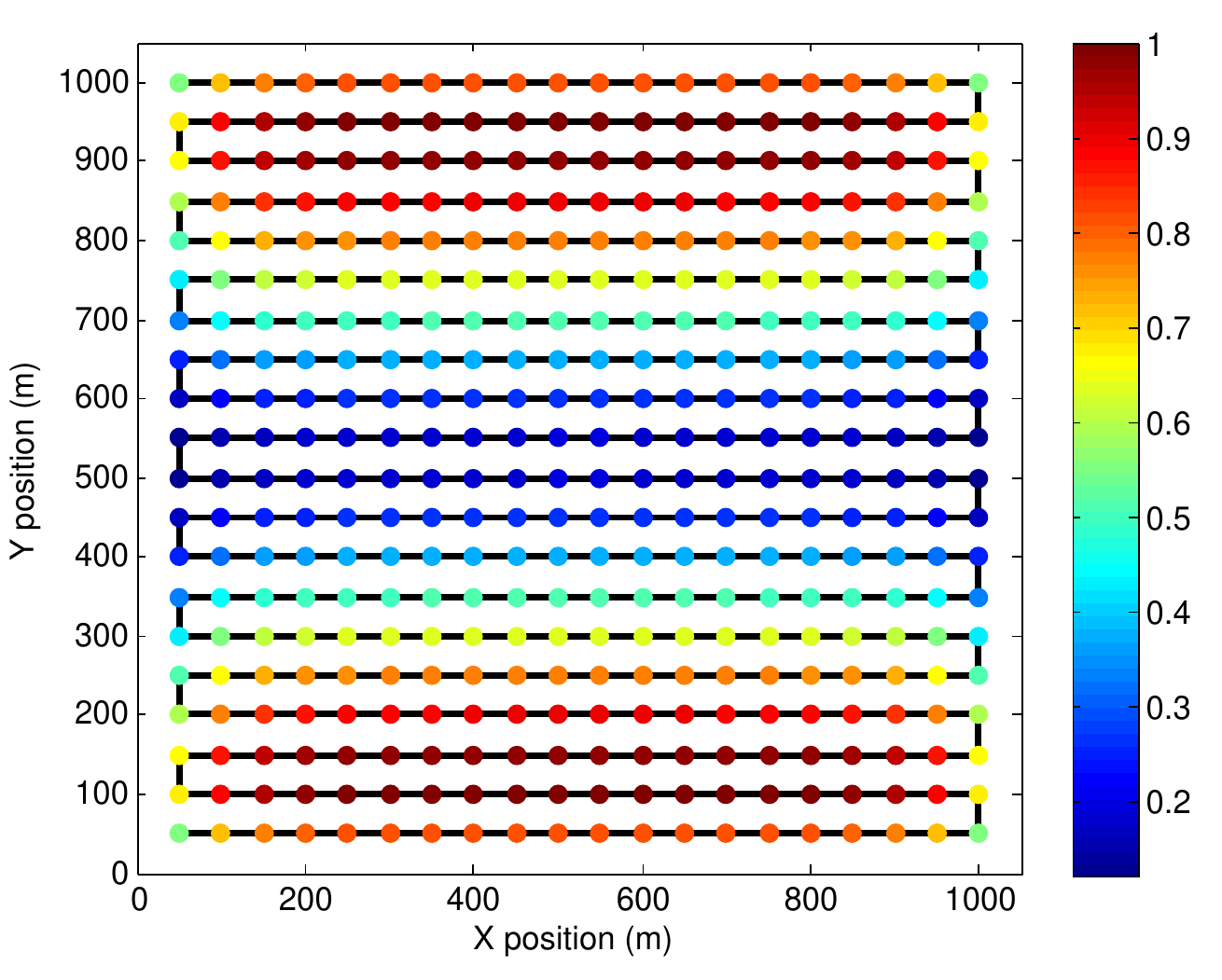}}
\\
\subfigure[]{\includegraphics[width=0.45\textwidth]{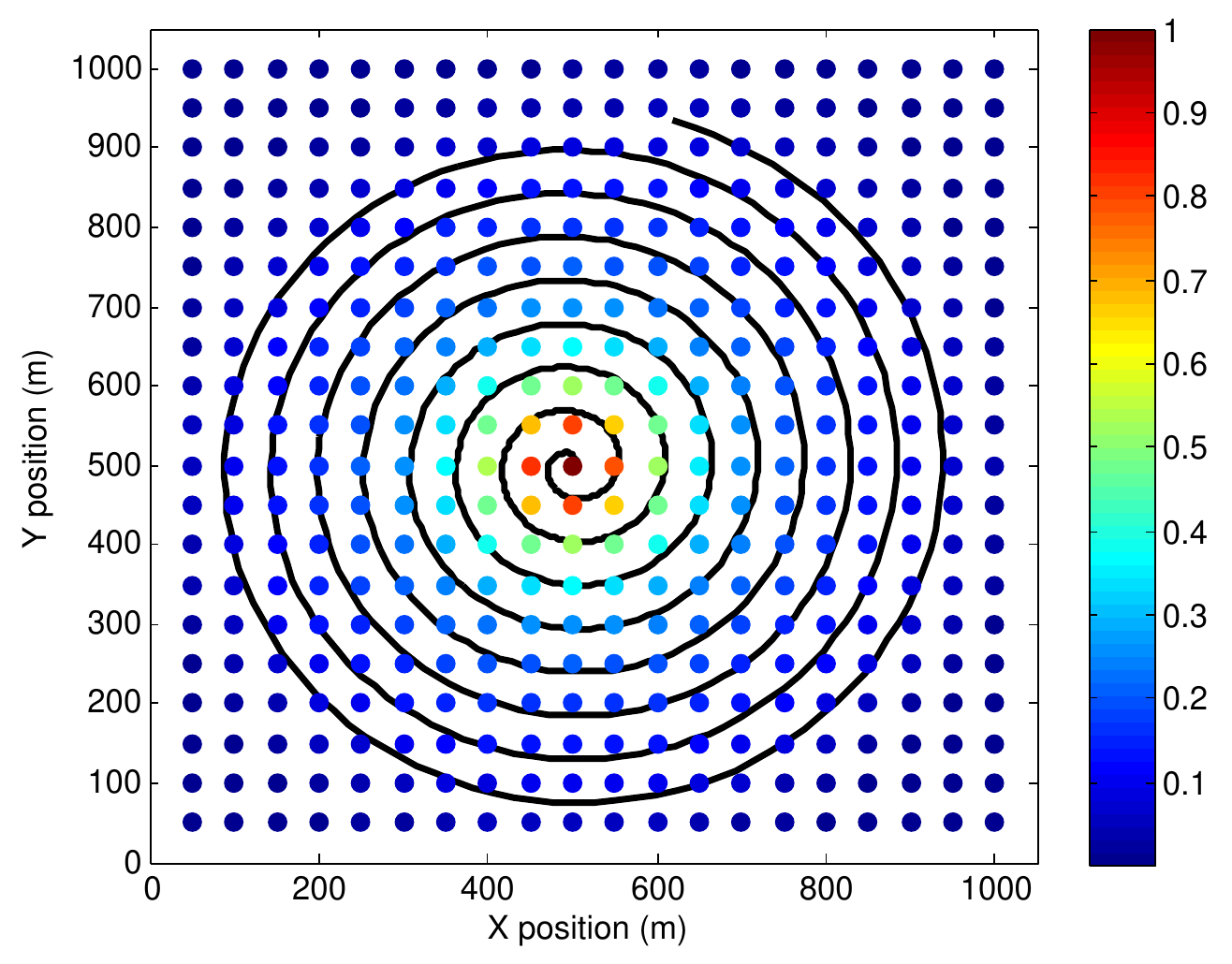}}
\subfigure[]{\includegraphics[width=0.45\textwidth]{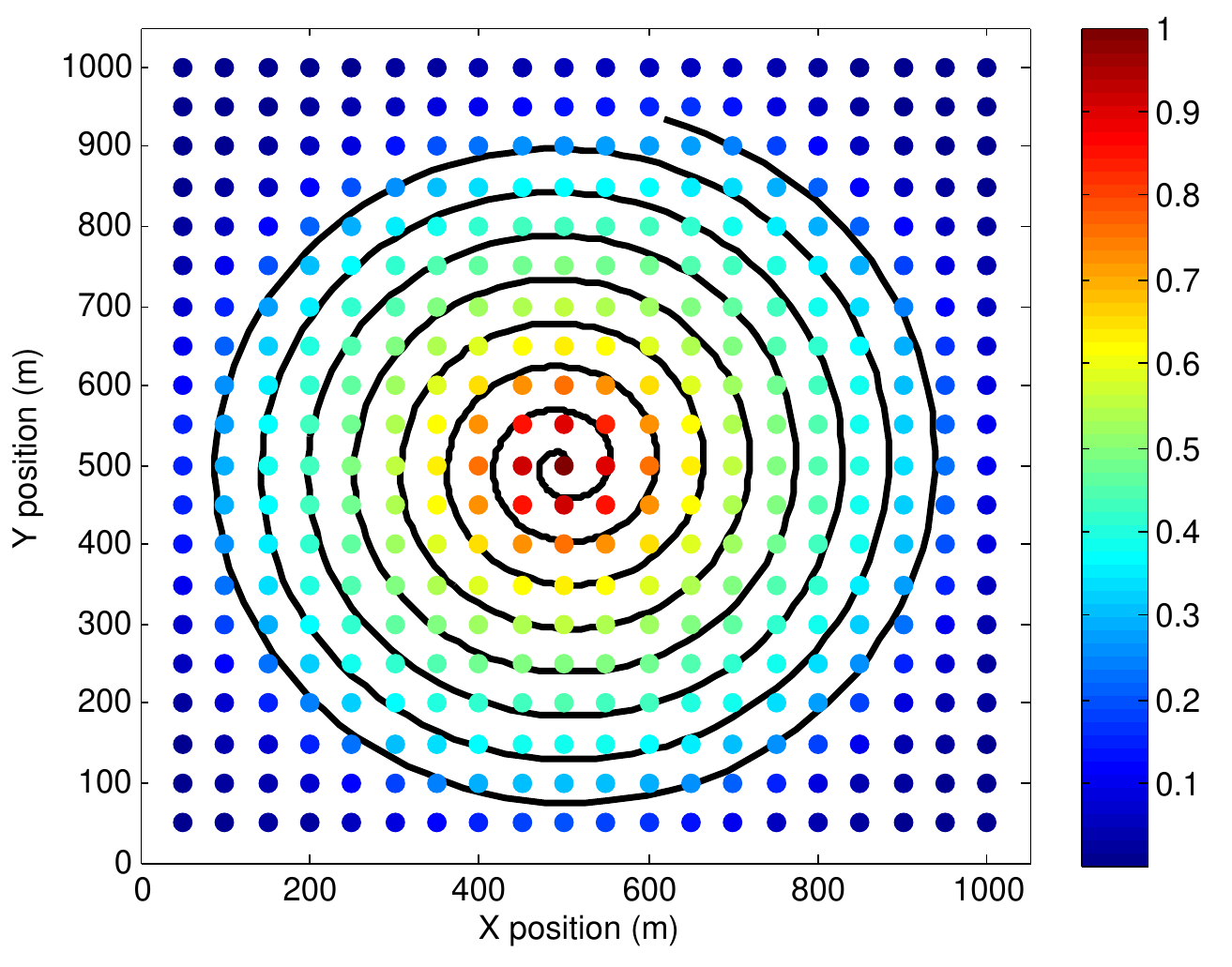}}
\caption{Normalized harvested energy under different trajectories with power control: a) sigmoid trajectory with fixed power; b) sigmoid trajectory with time varying power; c) spiral trajectory with fixed power; d) spiral trajectory with time varying power.}
\end{center}
\end{figure}

\newpage
\begin{figure}
\centering{\includegraphics[width=120mm]{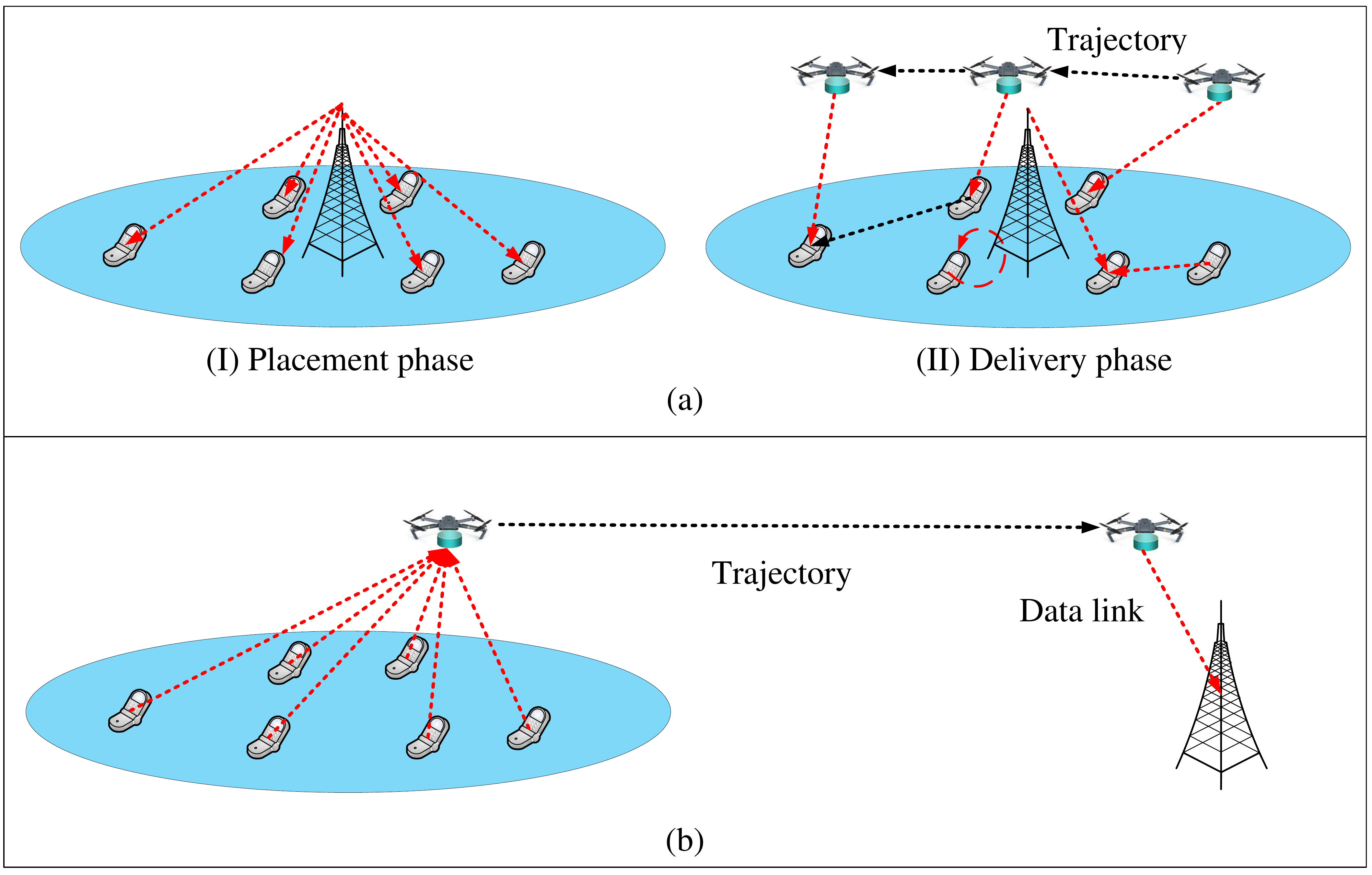}}
\caption{The UAV serves as mobile cache under two scenarios: a) serving mobile users; b) collecting and transmitting the information.}
\end{figure}
\end{document}